\RequirePackage[mathlines]{lineno}
\documentclass[aps,twocolumn,showpacs,preprintnumbers,amsmath,amssymb,superscriptaddress]{revtex4}

\usepackage{graphicx}
\usepackage{dcolumn}
\usepackage{bm}
\usepackage{color}
\usepackage{ulem}
\usepackage{url}
\usepackage{multirow}

\newcommand{ \bfg }{\begin{figure}[htpb]}
\newcommand{ \efg }{\end{figure}}
\newcommand{ \bmn }{\begin{minipage}}
\newcommand{ \emn }{\end{minipage}}
\newcommand{ \bt }{\begin{table}[htpb]}
\newcommand{ \et }{\end{table}}
\newcommand{ \pp }{$p$ + $p$ }
\newcommand{\dAu}{\textit{d} + Au }
\newcommand{\srt}{$\sqrt{s}$ }

\newcommand{\pt}{$p_T$ }

\newcommand{\Dzero}{$D^{0}$ }
\newcommand{\Dzerobar}{$\overline{D^{0}}$ }
\newcommand{\Dstar}{$D^{*}$ }

\newcommand{\dsigmady}{$d\sigma/dy|_{y=0}^{c\bar{c}}$}

\begin{document}
\def\Journal#1#2#3#4{{#1} {\bf #2}, #3 (#4)}

\def\NCA{Nuovo Cimento}
\def\NIM{Nucl. Instr. Meth.}
\def\NIMA{{Nucl. Instr. Meth.} A}
\def\NPB{{Nucl. Phys.} B}
\def\NPA{{Nucl. Phys.} A}
\def\PLB{{Phys. Lett.}  B}
\def\PRL{{Phys. Rev. Lett.}}
\def\PRC{{Phys. Rev.} C}
\def\PRD{{Phys. Rev.} D}
\def\ZPC{{Z. Phys.} C}
\def\JPG{{J. Phys.} G}
\def\JPGN{{J. Phys.} G: Nucl. Part. Phys.}
\def\EPJ{{Eur. Phys. J.} C}
\def\EPJST{{Eur. Phys. J.} - Special Topics}
\def\RPP{{Rep. Prog. Phys.}}
\def\JHEP{{J. High Energy Phys.}}

\preprint{}
\title{Measurements of \Dzero and \Dstar Production in \pp Collisions at \srt = 200 GeV}

\affiliation{AGH University of Science and Technology, Cracow, Poland}
\affiliation{Argonne National Laboratory, Argonne, Illinois 60439, USA}
\affiliation{Brookhaven National Laboratory, Upton, New York 11973, USA}
\affiliation{University of California, Berkeley, California 94720, USA}
\affiliation{University of California, Davis, California 95616, USA}
\affiliation{University of California, Los Angeles, California 90095, USA}
\affiliation{Universidade Estadual de Campinas, Sao Paulo, Brazil}
\affiliation{Central China Normal University (HZNU), Wuhan 430079, China}
\affiliation{University of Illinois at Chicago, Chicago, Illinois 60607, USA}
\affiliation{Creighton University, Omaha, Nebraska 68178, USA}
\affiliation{Czech Technical University in Prague, FNSPE, Prague, 115 19, Czech Republic}
\affiliation{Nuclear Physics Institute AS CR, 250 68 \v{R}e\v{z}/Prague, Czech Republic}
\affiliation{University of Frankfurt, Frankfurt, Germany}
\affiliation{Institute of Physics, Bhubaneswar 751005, India}
\affiliation{Indian Institute of Technology, Mumbai, India}
\affiliation{Indiana University, Bloomington, Indiana 47408, USA}
\affiliation{Alikhanov Institute for Theoretical and Experimental Physics, Moscow, Russia}
\affiliation{University of Jammu, Jammu 180001, India}
\affiliation{Joint Institute for Nuclear Research, Dubna, 141 980, Russia}
\affiliation{Kent State University, Kent, Ohio 44242, USA}
\affiliation{University of Kentucky, Lexington, Kentucky, 40506-0055, USA}
\affiliation{Institute of Modern Physics, Lanzhou, China}
\affiliation{Lawrence Berkeley National Laboratory, Berkeley, California 94720, USA}
\affiliation{Massachusetts Institute of Technology, Cambridge, MA 02139-4307, USA}
\affiliation{Max-Planck-Institut f\"ur Physik, Munich, Germany}
\affiliation{Michigan State University, East Lansing, Michigan 48824, USA}
\affiliation{Moscow Engineering Physics Institute, Moscow Russia}
\affiliation{Ohio State University, Columbus, Ohio 43210, USA}
\affiliation{Old Dominion University, Norfolk, VA, 23529, USA}
\affiliation{Panjab University, Chandigarh 160014, India}
\affiliation{Institute of Nuclear Physics PAS, Cracow, Poland}
\affiliation{Pennsylvania State University, University Park, Pennsylvania 16802, USA}
\affiliation{Institute of High Energy Physics, Protvino, Russia}
\affiliation{Purdue University, West Lafayette, Indiana 47907, USA}
\affiliation{Pusan National University, Pusan, Republic of Korea}
\affiliation{University of Rajasthan, Jaipur 302004, India}
\affiliation{Rice University, Houston, Texas 77251, USA}
\affiliation{Universidade de Sao Paulo, Sao Paulo, Brazil}
\affiliation{University of Science \& Technology of China, Hefei 230026, China}
\affiliation{Shandong University, Jinan, Shandong 250100, China}
\affiliation{Shanghai Institute of Applied Physics, Shanghai 201800, China}
\affiliation{SUBATECH, Nantes, France}
\affiliation{Texas A\&M University, College Station, Texas 77843, USA}
\affiliation{University of Texas, Austin, Texas 78712, USA}
\affiliation{University of Houston, Houston, TX, 77204, USA}
\affiliation{Tsinghua University, Beijing 100084, China}
\affiliation{United States Naval Academy, Annapolis, MD 21402, USA}
\affiliation{Valparaiso University, Valparaiso, Indiana 46383, USA}
\affiliation{Variable Energy Cyclotron Centre, Kolkata 700064, India}
\affiliation{Warsaw University of Technology, Warsaw, Poland}
\affiliation{University of Washington, Seattle, Washington 98195, USA}
\affiliation{Wayne State University, Detroit, Michigan 48201, USA}
\affiliation{Yale University, New Haven, Connecticut 06520, USA}
\affiliation{University of Zagreb, Zagreb, HR-10002, Croatia}

\author{L.~Adamczyk}\affiliation{AGH University of Science and Technology, Cracow, Poland}
\author{G.~Agakishiev}\affiliation{Joint Institute for Nuclear Research, Dubna, 141 980, Russia}
\author{M.~M.~Aggarwal}\affiliation{Panjab University, Chandigarh 160014, India}
\author{Z.~Ahammed}\affiliation{Variable Energy Cyclotron Centre, Kolkata 700064, India}
\author{A.~V.~Alakhverdyants}\affiliation{Joint Institute for Nuclear Research, Dubna, 141 980, Russia}
\author{I.~Alekseev}\affiliation{Alikhanov Institute for Theoretical and Experimental Physics, Moscow, Russia}
\author{J.~Alford}\affiliation{Kent State University, Kent, Ohio 44242, USA}
\author{B.~D.~Anderson}\affiliation{Kent State University, Kent, Ohio 44242, USA}
\author{C.~D.~Anson}\affiliation{Ohio State University, Columbus, Ohio 43210, USA}
\author{D.~Arkhipkin}\affiliation{Brookhaven National Laboratory, Upton, New York 11973, USA}
% ...

\collaboration{STAR Collaboration}\noaffiliation


\date{\today}
\begin{abstract}
We report measurements of charmed-hadron ($D^{0}$, $D^{*}$) production cross sections at mid-rapidity in \pp collisions at a center-of-mass energy of 200\,GeV by the STAR experiment. Charmed hadrons were reconstructed via the hadronic decays $D^{0}\rightarrow K^{-}\pi^{+}$, $D^{*+}\rightarrow D^{0}\pi^{+}\rightarrow  K^{-}\pi^{+}\pi^{+}$ and their charge conjugates, covering the $p_T$ range of 0.6$-$2.0 GeV/$c$ and 2.0$-$6.0 GeV/$c$ for  $D^{0}$ and $D^{*+}$, respectively. From this analysis, the charm-pair production cross section at mid-rapidity is \dsigmady = 170 $\pm$ 45 (stat.) $^{+38}_{-59}$ (sys.) $\mu$b. The extracted charm-pair cross section is compared to perturbative QCD calculations. The transverse momentum diﬀerential cross section is found to be consistent with the upper bound of a Fixed-Order Next-to-Leading Logarithm calculation.  

\end{abstract}
\maketitle


\section{Introduction}
The primary goal of ultra-relativistic heavy-ion experiments at the
Relativistic Heavy Ion Collider (RHIC) is to search for and
characterize the new state of matter with partonic degrees of freedom,
namely the Quark Gluon Plasma (QGP), predicted by Quantum
ChromoDynamics (QCD)~\cite{STARwhitepaper}. In high-energy collisions at RHIC, heavy quarks ($c$, $b$)
are expected to be created from initial hard scatterings~\cite{LinPRC} and the relative changes in their masses are small by the strong interactions with the QCD
medium~\cite{quarkMass}. Thus they carry clean information from the
system at the early stage. The interaction between heavy quarks and the
medium is sensitive to the medium dynamics, therefore heavy quarks are
suggested as an ``ideal'' probe to quantify the properties of the
strongly interacting QCD matter~\cite{MooreTeaney,vanHeesRapp,Uphoff}. Consequently, measurements of heavy-quark production over a wide transverse momentum ($p_T$) region in proton-proton ($p$ + $p$) collisions are critical to provide a baseline for understanding the results from heavy-ion collisions. In particular, precise knowledge of the total charm production cross sections from \pp to central heavy-ion collisions is critical to understand both open charm and charmonium production mechanisms in the QGP medium formed in central heavy-ion collisions at RHIC~\cite{Thews,PBMstat}. 

In elementary particle collisions, processes involving heavy quarks
with masses much larger than the QCD scale ($\Lambda_{\text{QCD}}$)
are, in principle, amenable to perturbative QCD (pQCD)
calculations. For heavy-quark production cross sections at large momentum transfer $Q^2$,
Fixed-Order Next-to-Leading Logarithm (FONLL) pQCD calculations, where $p_{T} \gg m_{c}$, are expected to work
reasonably well~\cite{FONLL}. However, calculations of the charm cross
section at low $p_T$ become complicated because charm quarks cannot be
treated as a massless flavor. Furthermore, in the low momentum
transfer region there is a large uncertainty in the gluon
density function, and the strong coupling constant increases dramatically. Thus, perturbative QCD calculations have little predictive power for the total charm cross section in high-energy hadron-hadron collisions~\cite{vogtTotalCharm}. In view of these theoretical issues, experimental measurements become necessary and in turn provide constraints that improve theoretical calculations.

Measurements of inclusive charm production have been carried out through two main approaches: \textit{i}) single leptons from heavy flavor semi-leptonic decays, and \textit{ii}) charmed hadrons from hadronic decays. The advantages of the first method include an experimentally triggerable observable and relatively large decay branching ratios, thus resulting in relatively large statistics. However, interpretations of the experimental results contain ambiguities because \textit{a}) leptons are produced by various charmed and bottomed hadron decays, and \textit{b}) heavy-flavor hadrons contributing to leptons at a certain \pt can come from a wide kinematic region due to the decay smearing. The second method suffers from a large combinatorial background when all particles from the collision vertex are included, without any reconstruction of the secondary weak-decay vertices. This background is particularly large (S/B is in the order of $1:10^3$) in heavy-ion collisions. 

There are many measurements of the charm production cross section in low energy \pp or $p$ + A collisions via both semi-leptonic and hadronic decays at CERN and Fermilab~\cite{lowEcharmReview,lowEcharm}. Results for the total charm cross sections (from measurements with reasonable extrapolations) are consistent with Next-to-Leading-Order (NLO) pQCD calculations. At high energies, the Collider Detector at Fermilab (CDF) collaboration at the Tevatron measured the charmed-hadron cross sections at $p_T>$ 5 GeV/$c$ in $p$ + $\bar{p}$ collisions at \srt = 1.96 TeV, and results for \Dzero, $D^{+}$ and $D^{*+}$ mesons are consistent with the upper bounds of FONLL pQCD calculations~\cite{CDFcharm}. 
At RHIC energies, charm production has been studied mainly via semi-leptonic decay electrons from \pp to Au + Au collisions~\cite{dAuCharm,STARnpe,STARnpepp,phenixNPEpp,phenixNPEAuAu}. The result from \pp collisions is also consistent with the upper bound of FONLL pQCD calculations at $p_T(e)>2$ GeV/$c$. Measurements of the \Dzero cross section by the reconstruction of hadronic decays were carried out in $d$ + Au collisions~\cite{dAuCharm}, but no measurement of the charmed-hadron production cross section in \pp collisions has been made at RHIC until now.

In this paper, we report measurements from the STAR experiment of the charmed-hadron ($D^{0}$, $D^{*}$) production cross section at mid-rapidity in \pp collisions at \srt = 200\,GeV. Charmed hadrons,  \Dzero and $D^{*}$, were reconstructed via hadronic decays in the transverse momentum ranges of 0.6$-$2.0 GeV/$c$ and 2$-$6 GeV/$c$, respectively. The $p_T$ differential production cross sections are compared to pQCD theoretical calculations, and a total charm cross section is extracted.

The paper is organized as follows: Section II describes the experimental setup, the data set, and the particle identification method used in this analysis. Section III explains the hadronic reconstruction for \Dzero and \Dstar mesons in detail. Section IV discusses the reconstruction efficiency, acceptance, and trigger/vertex corrections. Details of the systematic uncertainties are discussed in Section V. The transverse momentum differential production cross section is presented in Section VI and it is compared with pQCD FONLL and PYTHIA~\cite{pythia} calculations. The results are summarized in Section VII.

\section{Experimental Setup}
\subsection{Detector Apparatus}
The data used in this analysis were recorded by the Solenoidal Tracker At RHIC (STAR) detector~\cite{starnim}. The STAR detector is a multi-purpose spectrometer with large rapidity coverage. 
The major subsystems at mid-rapidity sit inside a solenoidal magnet which provides a uniform magnetic field of 0.5 Tesla along the beam axis. Subsystems used in this analysis are the Time Projection Chamber (TPC)~\cite{TPCnim}, the Time-Of-Flight (TOF) detector~\cite{tofproposal}, the barrel and endcap Electromagnetic Calorimeters (EMC)~\cite{bemcnim,eemcnim}, and two trigger detector subsystems: the Vertex Position Detector (VPD)~\cite{VPDnim} and the Beam Beam Counters (BBC)~\cite{bbcnim}.

The TPC is the main tracking detector, covering the full azimuthal
angle at pseudo-rapidity $|\eta|<$ 1 for tracks crossing all 45
padrows~\cite{TPCnim}. It measures the charged-particle momenta and
provides particle-identification (PID) capability via the ionization energy loss ($dE/dx$) in
the TPC gas, allowing a clean separation between charged kaons and
pions up to momentum $p\sim$0.6 GeV/$c$. The barrel TOF detector is a newly installed subsystem, utilizing the multi-gap resistive plate chamber technology~\cite{tofproposal}. The full system consists of 120 trays covering the full azimuth at $|\eta|<$ 0.9 surrounding the TPC cylinder. In the year 2009 run, 84 trays out of 120 for the full barrel were installed and used for this analysis. The TOF detector uses the timing recorded in the forward VPD detector as the start time to calculate the particle time of flight, which is combined with the momentum from the TPC to identify particles. The timing resolution of the TOF system, including the start timing resolution in \srt = 200\,GeV \pp collisions, is about 110 ps, allowing separation of $K$ and $\pi$ up to $p\sim$1.5 GeV/$c$. The barrel and endcap EMCs are designed to identify electrons and photons, covering the full azimuthal angle at $|\eta|<$ 1 and 1 $<\eta<$ 2, respectively~\cite{bemcnim,eemcnim}. They are fast-response detectors ($< 100$ ns), and were used to suppress the TPC pileup-track contribution in the event-vertex finder by matching with charged tracks from the TPC.

In addition to providing the start time for the barrel TOF detector, the VPD detector is also one of the trigger detectors in STAR. It has two parts surrounding the beam pipe, located on the east and west sides, 5.7 m away from the center of the STAR detector and covering 4.24 $<|\eta|<$ 5.1~\cite{VPDnim}. The minimum-bias (MB) trigger was defined as a coincidence signal in the east and west VPD detectors and a selection was made on the vertex position along the beam axis ($V_z$) to be within 40 cm of the center of the STAR detector. The BBC~\cite{bbcnim} consists of two identical counters located on each side of the TPC covering full azimuth and 2.1 $<|\eta|<$ 5.0 in pseudo-rapidity. Each part consists of a set of hexagonal scintillator tiles grouped into a ring and mounted around the beam pipe at a distance of 3.7 m from the center of STAR. The BBC detector had been used to define the main minimum-bias trigger in \pp collisions before the minimum-bias trigger was used in 2009. A small sample of BBC minimum-bias-triggered events were collected in 2009 to check for a trigger bias. Details of the minimum-bias trigger bias and correction will be discussed in Sect. IV.

\subsection{Data Sets and Event Selection}
The data sample used in this analysis consisted of minimum-bias-triggered \pp collisions at $\sqrt{s} = 200$ GeV,  recorded in 2009 by the STAR experiment at RHIC. 

\begin{figure}[ht]
\centerline{
\includegraphics[width=0.45\textwidth] {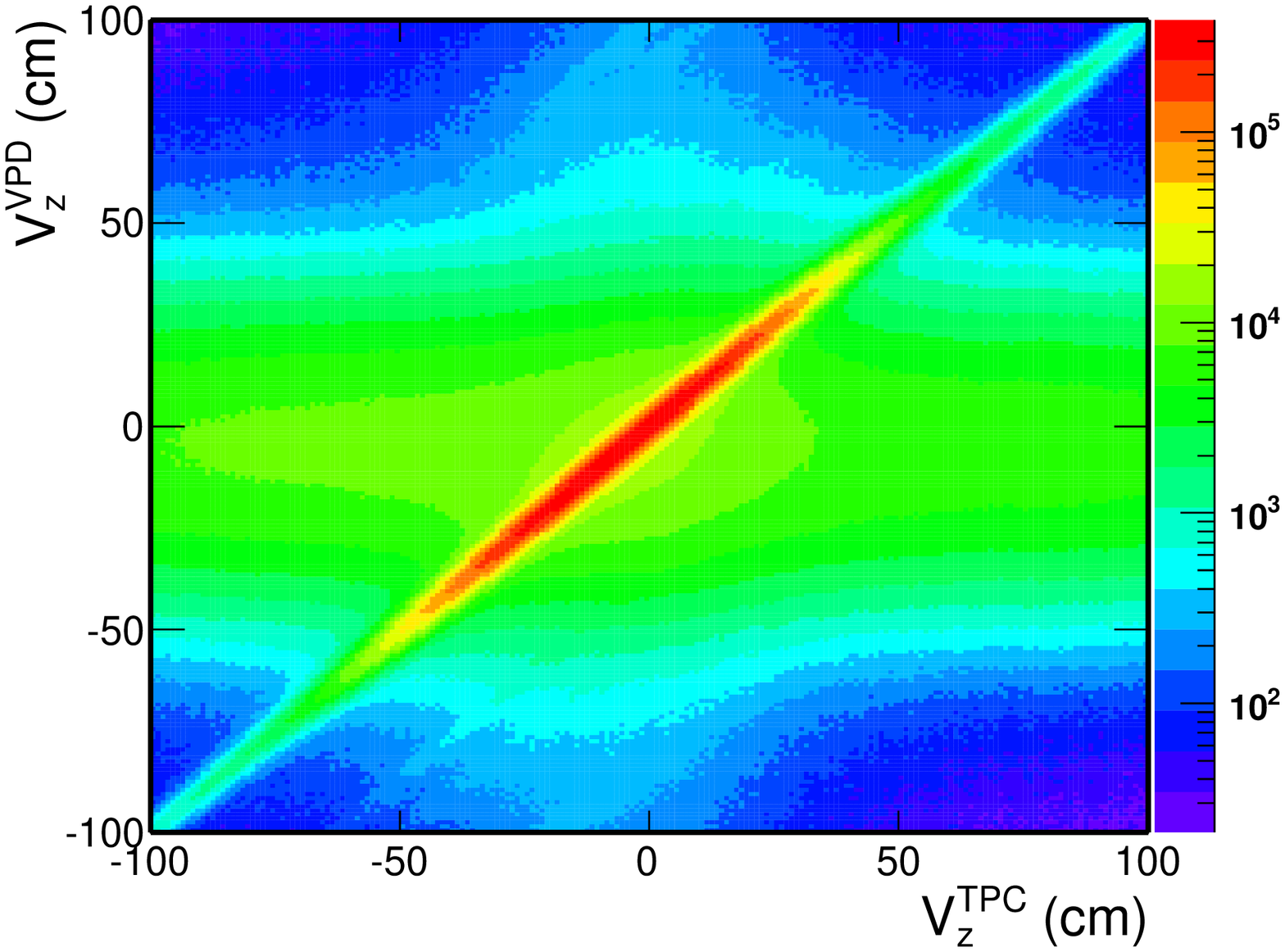}}
\centerline{
\includegraphics[width=0.45\textwidth] {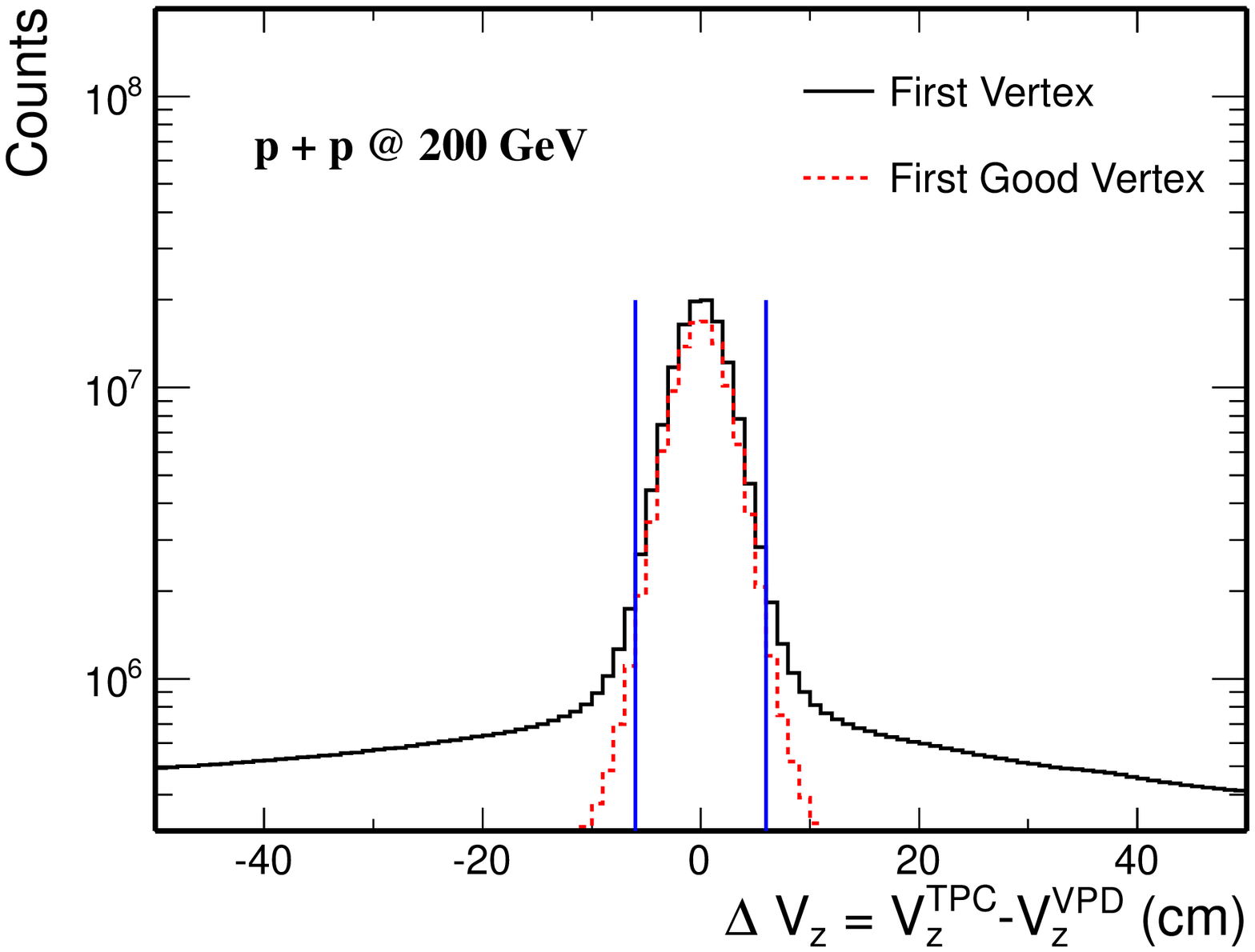}}
 \caption[]{(color online) Upper Panel: Correlation of $V^{TPC}_{z}$ versus
   $V^{VPD}_{z}$. Bottom Panel: $\Delta V_z$ distributions. A ``good'' vertex requirement rejects most of the pileup events. Blue vertical lines indicate the cuts for the $V_z$ selection.}
 \label{fig1vz}
\end{figure}

The intrinsic drift time for electrons from the center to one end of the TPC is on the order of 40 $\mu s$. Thus, in high-luminosity \pp collisions, one TPC event usually contains tracks from collisions originating from non-triggered bunch crossings. These ``pileup events'' will lead to additional tracks recorded in the TPC, in addition to those from the triggered event. This effect was not significant in previous RHIC runs, but the increase in the collision rate during 2009 to several hundred kHz made this a significant effect. 
The $V_z$ position from offline VPD data 
has a resolution of 2.5 cm for minimum-bias events, which can provide
a useful constraint to select the real event that fired the
trigger. Figure~\ref{fig1vz}, upper panel, shows the correlation
between the $V_z$ positions from the TPC and the VPD. 
Events with TPC vertices along the diagonal correlated band are real
ones that fired the VPD minimum bias trigger.
In Fig.~\ref{fig1vz}, bottom panel, the solid black histogram shows the 1-D $V_z$ difference between the first TPC-determined vertex position and VPD-determined vertex position. By applying a $V_z$ difference cut $|\Delta Vz|<$ 6 cm, most of the TPC pileup events can be removed. There still remain random associated correlations that enter into this cut window ($\sim$7\% level, calculated using a two-Gaussian fit). To further suppress this contamination, we required the TPC event vertices to have at least two tracks that match with hits in the barrel and endcap EMCs (this vertex is treated as a ``good'' vertex). The red dashed histogram in Fig.~\ref{fig1vz}, bottom panel, shows the $\Delta Vz$ distribution after this selection. The random associated pileup events in the $V_z$ difference cut window are now suppressed to $\sim$2\% of the total, while the corresponding loss of real events is $\sim$15\%. In total, 105 million minimum-bias events were used in the charmed-hadron analysis.

\subsection{Track Reconstruction and Particle Identification}

Charged particle tracks are required to point within $|\eta|<$ 1 in
order minimize TPC acceptance effects during reconstruction. Tracks
must have 15 out of a maximum of 45 points used in track fitting
(nFitPts), and at least 52\% of the total possible fit points in order to avoid double-counting split tracks. Tracks are required to have a distance-of-closest-approach (DCA) to the collision vertex of less than 2 cm to suppress background tracks produced by secondary scattering in the detector and also long-lived particle decays. The STAR track pointing resolution with the TPC alone does not have the precision to separate charm secondary decay vertices from the collision vertices. 

\begin{figure}[ht]
\centerline{
\includegraphics[width=0.45\textwidth] {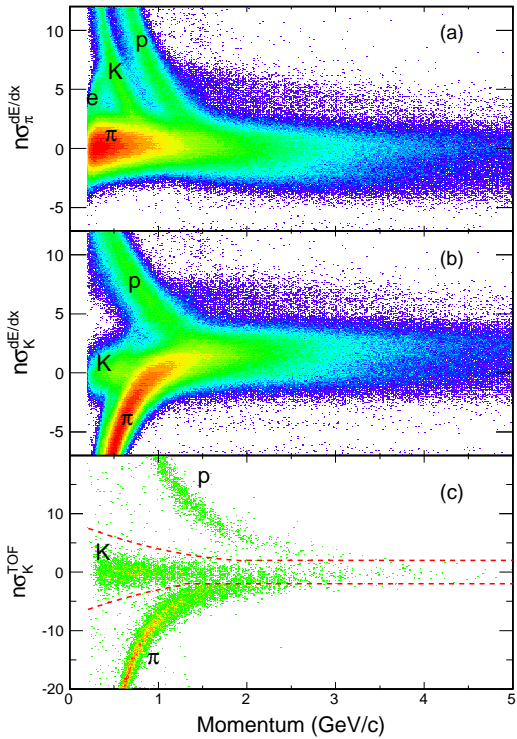}}
\caption[]{(color online) Distributions of n$\sigma_{\pi}^{dE/dx}$, n$\sigma_{K}^{dE/dx}$, and n$\sigma_{K}^{TOF}$ versus momentum are shown in panels (a), (b), and (c), respectively. The latter is shown after $dE/dx$ cuts were applied.}
\label{fig2PID}
\end{figure}

Particle identification for final-state charged hadrons was carried out with a combination of $dE/dx$ in the TPC and the particle velocity ($\beta$) measurement from the barrel TOF detector. Thus the normalized $dE/dx$ ($n\sigma_{X}^{dE/dx}$) and $1/\beta$ ($n\sigma_{X}^{\text{TOF}}$) distributions were used to select daughter particle candidates. They are defined as follows:

\begin{eqnarray}
n\sigma_{X}^{dE/dx} & = & \frac{\text{ln}\frac{\langle dE/dx \rangle^{\text{mea}}}{dE/dx^{\text{th}}_X}}{R_{dE/dx}} \\
n\sigma_{X}^{\text{TOF}} & = &\frac{\frac{1}{\beta^{\text{mea}}}-\frac{1}{\beta^{\text{th}}_X}}{R_{1/\beta}}
\end{eqnarray}

where the superscripts ``$\text{mea}$'' and ``$\text{th}$'' are measured and theoretical values, respectively. The $X$ denotes expected values which are calculated with respect to one kind of particle species ($\pi$ or $K$). $R_{dE/dx}$ and $R_{1/\beta}$ are the experimental $dE/dx$ and $1/\beta$ resolutions, respectively. With the above definitions, the two resulting distributions can be approximated by Gaussian distributions with mean$\sim$0 and $\sigma\sim$1). Figure~\ref{fig2PID} shows the $n\sigma_K^{dE/dx}$, $n\sigma_{\pi}^{dE/dx}$, and $n\sigma_K^{\text{TOF}}$ distributions versus particle momentum. 

Daughter kaon (pion) candidates are selected by requiring  $|n\sigma_K^{dE/dx}|<$ 2 ($|n\sigma_\pi^{dE/dx}|<$ 2). In addition, to improve the significance of the reconstructed \Dzero signal, the kaon daughter tracks were required to have a valid hit in the TOF detector and then selected with a TOF PID cut, which is denoted as the red dashed lines in panel (c) of Fig.~\ref{fig2PID}. In order to have good efficiency and considering pion identification is good enough with $dE/dx$ only, we did not require pion to match with TOF.

\section{Charmed-hadron reconstruction and raw yield extraction}
\subsection{\Dzero Reconstruction}
\begin{figure}[htbp]
\centerline{\includegraphics[width=0.48\textwidth]{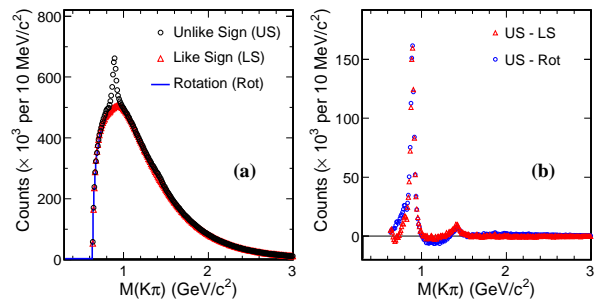}} 
\caption[]{(color online) (a) Invariant mass distributions of raw $K\pi$ combinations for unlike-sign pairs (circles), like-sign pairs (triangles), and kaon momentum rotated pairs (line). (b) Residual distributions after subtracting the like-sign distribution (triangles) and rotation pair distribution (dots) from the unlike-sign distribution. } 
\label{fig3d0total}
\end{figure}

\begin{figure}[floatfix] 
\centerline{\includegraphics[width=0.48\textwidth]{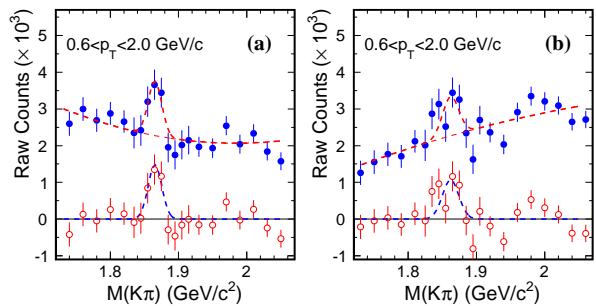}} 
\caption[]{(color online) Invariant $K\pi$ mass distributions in the \Dzero mass region after like-sign (a) and track-rotation (b) background subtraction. Solid circles show the signal and a residual background. A Gaussian function and a 2nd order polynomial function were used to describe the signal and residual background, respectively. Open circles show the signal after residual background subtraction.} 
\label{fig4d0signal}
\end{figure}

\Dzero and \Dzerobar mesons were reconstructed via the hadronic decay
$D^{0}(\overline{D^{0}}) \rightarrow K^{\mp}\pi^{\pm}$ with a
branching ratio of 3.89\%. The analysis technique is the same as that
used for a \Dzero analysis in \dAu collisions~\cite{dAuCharm}. In \pp
collisions, the mixed-events technique is not suitable for describing
the background due to large contribution of correlated
jets. Therefore, two different techniques were used to reproduce the
background: the like-sign and track-rotation methods. Since the $\pi^{-}$ and $\pi^{+}$ production is symmetric in the STAR uniform acceptance and their yield ratio is measured to be 0.988 $\pm$ 0.043~\cite{SysPID}, the like-sign
(LS) method is used and a pair combination with the same charged
sign is expected to reproduce the background without the signal
correlation. The opposite-sign backgrounds, which go into the residual background, are only several percent of total background and will be discussed later. The track-rotation (Rot) technique has been used in many measurements~\cite{Rotref}. This method is based on the
assumption that by rotating the daughter kaon track by 180 degrees in
azimuth, the decay kinematics are destroyed. Thus the invariant mass
distribution after rotation is able to reproduce the random
combinatorial background. Figure~\ref{fig3d0total} shows the invariant
mass distributions of $K\pi$ candidates. Panel (a) shows the invariant
mass distributions for $K\pi$ pairs ($0.6 < p_{T}(K\pi) < 2.0$
GeV/$c$) with unlike-sign (US) before background subtraction, with
like-sign, and with rotated kaon momentum. The distributions from the
like-sign and track-rotation techniques describe the background
well. Panel (b) is the unlike-sign $K\pi$ invariant mass distribution
after combinatorial background subtraction. A significant $K^*$(892)
peak is observed. The secondary small peak at about 1.4 GeV/$c^2$ is
the $K^*_{2}$(1430).  A direct zoom-in view of the vicinity around the
\Dzero mass region is shown in 
Fig.~\ref{fig4d0signal} (panel (a) for subtraction of like-sign
background, and panel (b) for the rotational case). Solid symbols
depict the same distributions as shown in Fig.~\ref{fig3d0total} and Fig.~\ref{fig5d0signalpT} in
two different \Dzero $p_T$ bins. One
can see there is still some ``residual'' background after like-sign or
rotational background subtraction. The possible sources
to the residual background have been investigated using PYTHIA
simulations. We performed the same reconstruction as we did on the
data, for the foreground and background distributions. From these simulations, we have learned
that the possible sources that can contribute to this
residual correlated background include: correlated hadron pairs from decays (mostly
resonances) where the real daughters were mis-identified as $K\pi$
pairs; $K\pi$ pair from other decay channels of \Dzero (e.g. $K^-\pi^+\pi^{0}$) where the other
daughters are missed in the reconstruction; same-charge $K^-\pi^-$ pairs from multi-body decays of $D^0 \rightarrow K^{-}\pi^{+}\pi^{+}\pi^{-}$; 
$K\pi$ pairs from jet fragmentations; etc. The different shape of the residual background from LS and Rot background subtraction in the data can be qualitatively reproduced by PYTHIA simulation. The magnitude of the residual background depends on how to choose the normalization for the
like-sign or rotational background, as qualitatively understood
from the PYTHIA simulations. However, the change of the residual
background magnitude due to different normalizations has a very small
impact on the final extracted signal counts, and it has been included
in the systematic uncertainties. We used an empirical
polynomial function to describe it and the choice of this empirical
function was also included as one of the systematic source to the raw
yields. 
A Gaussian function is used to fit the signal. The raw yield of the \Dzero is obtained by fitting the data (blue solid circles) with a fit function representing the sum of signal and background (red dashed curve) in the mass region of $1.72 < M_{K\pi} < 2.05$ GeV/$c^2$. The signal after the residual background subtraction is shown as the red open circles. The Gaussian function used to describe the signal is shown as the blue dashed curve. The total \Dzero signal consists of 4085 $\pm$ 938 counts.

The signals after background subtraction for two \pt bins are shown in
Fig.~\ref{fig5d0signalpT}. Panels (a), (c) and (b), (d) show the signals from LS
and Rot background subtraction, respectively. The \Dzero raw yields and statistical errors extracted from the two background methods are listed in Table~\ref{D0yields}. The average values of
the \Dzero counts from the LS and Rot background methods are used to
calculate the final \Dzero raw yield in each \pt bin. The mean and
width from the Gaussian fits are compared with MC simulation in
Fig.~\ref{meanwidth} (left panels). The single \Dzero and \Dstar are embedded into the real data and simulated in the full STAR GEANT reconstruction chain, taking into account detector response and material effect. The \Dzero signal mean value from an
open-parameter fit shifts to lower mass due to kaon energy loss at low
$p_T$, which is not fully accounted in the simulation due to possibly
missing material budget. The
systematic uncertainty in determining the \Dzero raw yields as well as
the potential double-counting issue due to particle misidentification will be discussed in Sect. V A.

\begin{table}[htbp]
\caption[]{$D^{0}$ raw yields.} \label{D0yields} \vskip 0.1 in
\centering\begin{tabular}{|c|c|c|} \hline
$p_T$ range (GeV/$c$) & 0.6$-$1.2 & 1.2$-$2 \\ \hline
$p_T$ (GeV/$c$) & 0.908 & 1.57 \\ \hline
raw yields $\times 10^3$ (Rot) & 2.45 $\pm$ 0.66 & 1.65 $\pm$ 0.63  \\ \hline
raw yields $\times 10^3$ (LS) & 1.67 $\pm$ 0.74 & 2.40 $\pm$ 0.64 \\
\hline
\end{tabular}
\end{table}

\bfg \centering
\includegraphics[width=0.48\textwidth]{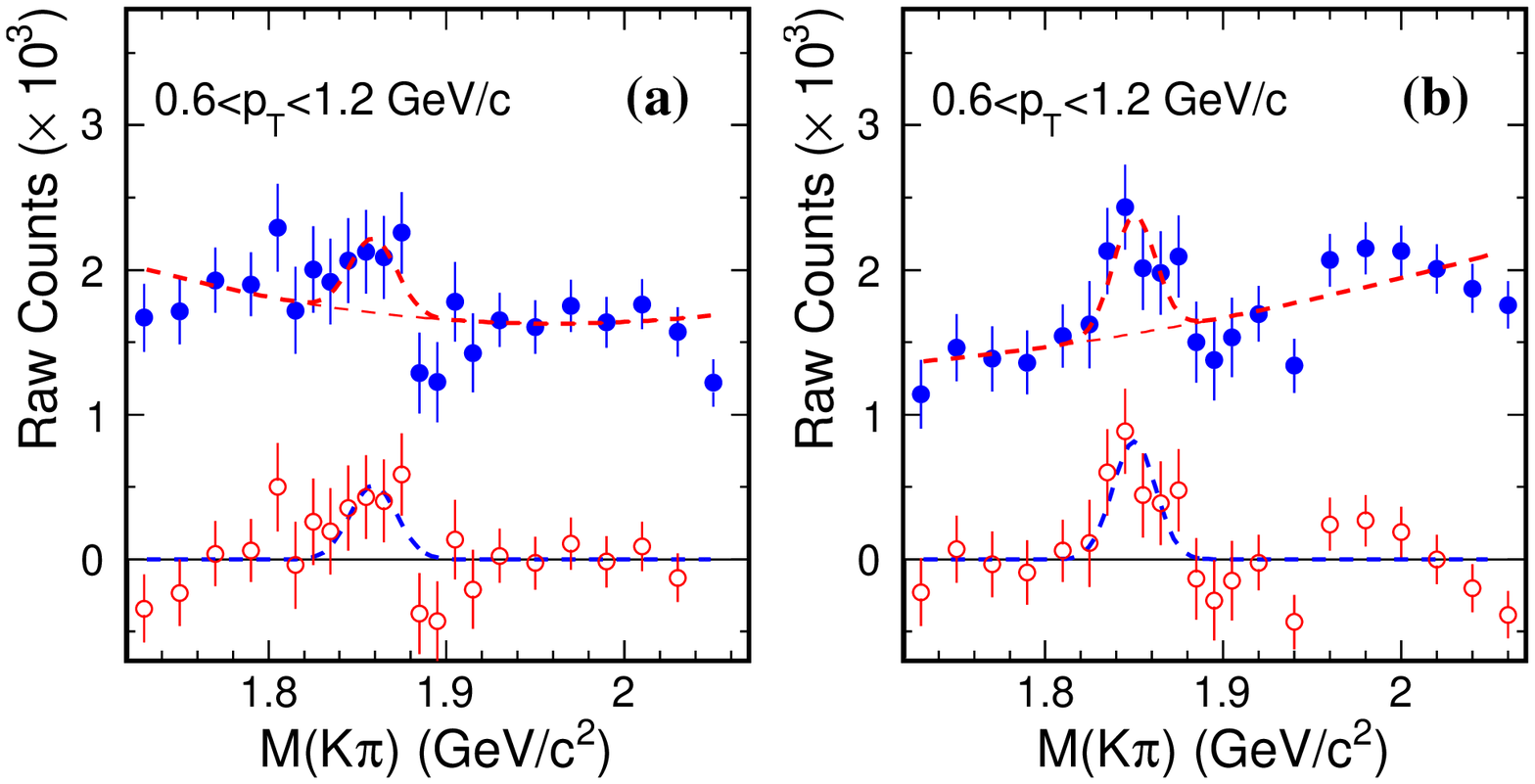} \\
\includegraphics[width=0.48\textwidth]{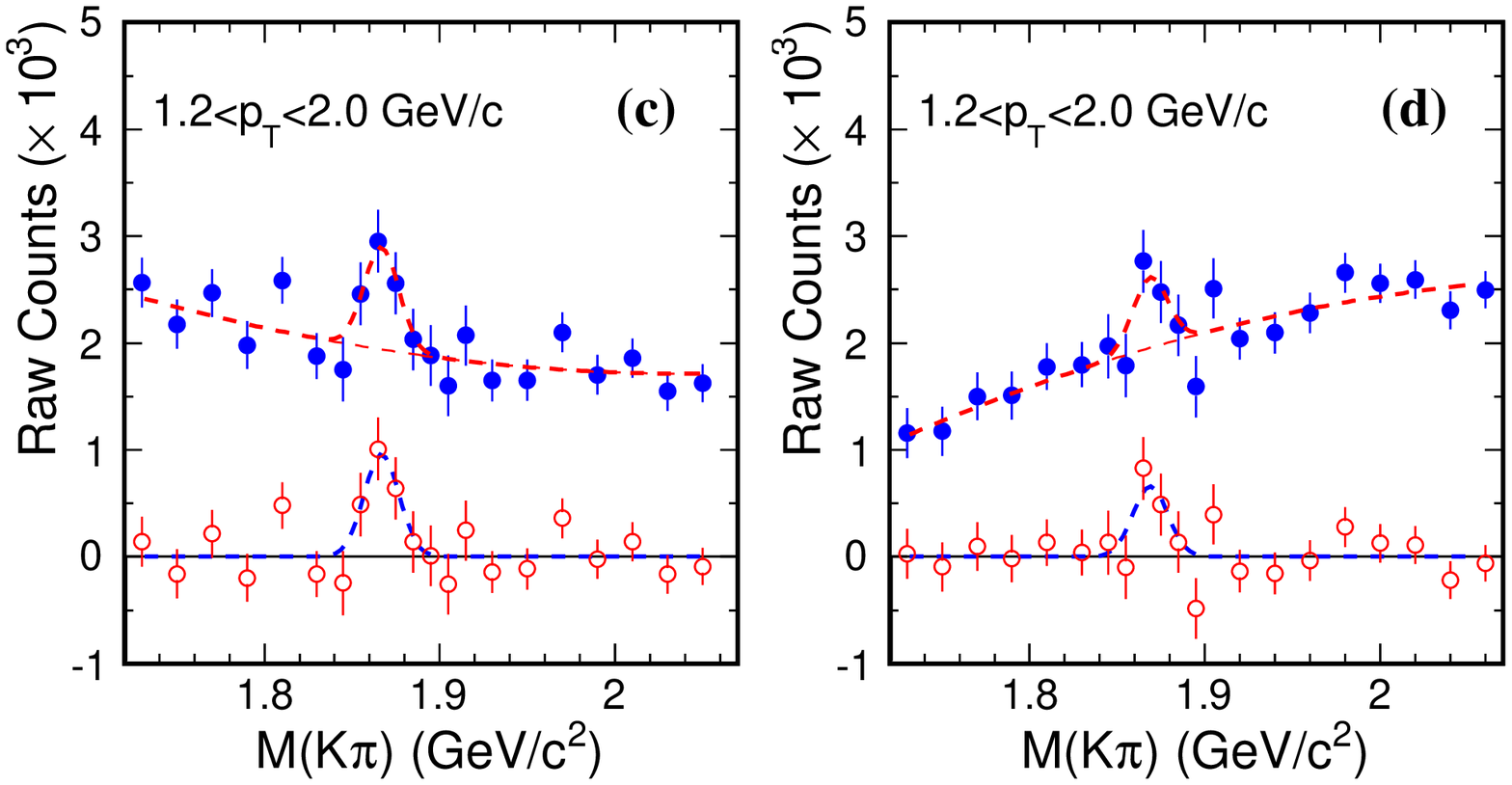} \\
\caption{(color online) Raw \Dzero signals in different $p_T$ bins after like-sign (a)(c) and track-rotation (b)(d) subtraction.}
\label{fig5d0signalpT} \efg

\bfg \centering
\includegraphics[width=0.5\textwidth]{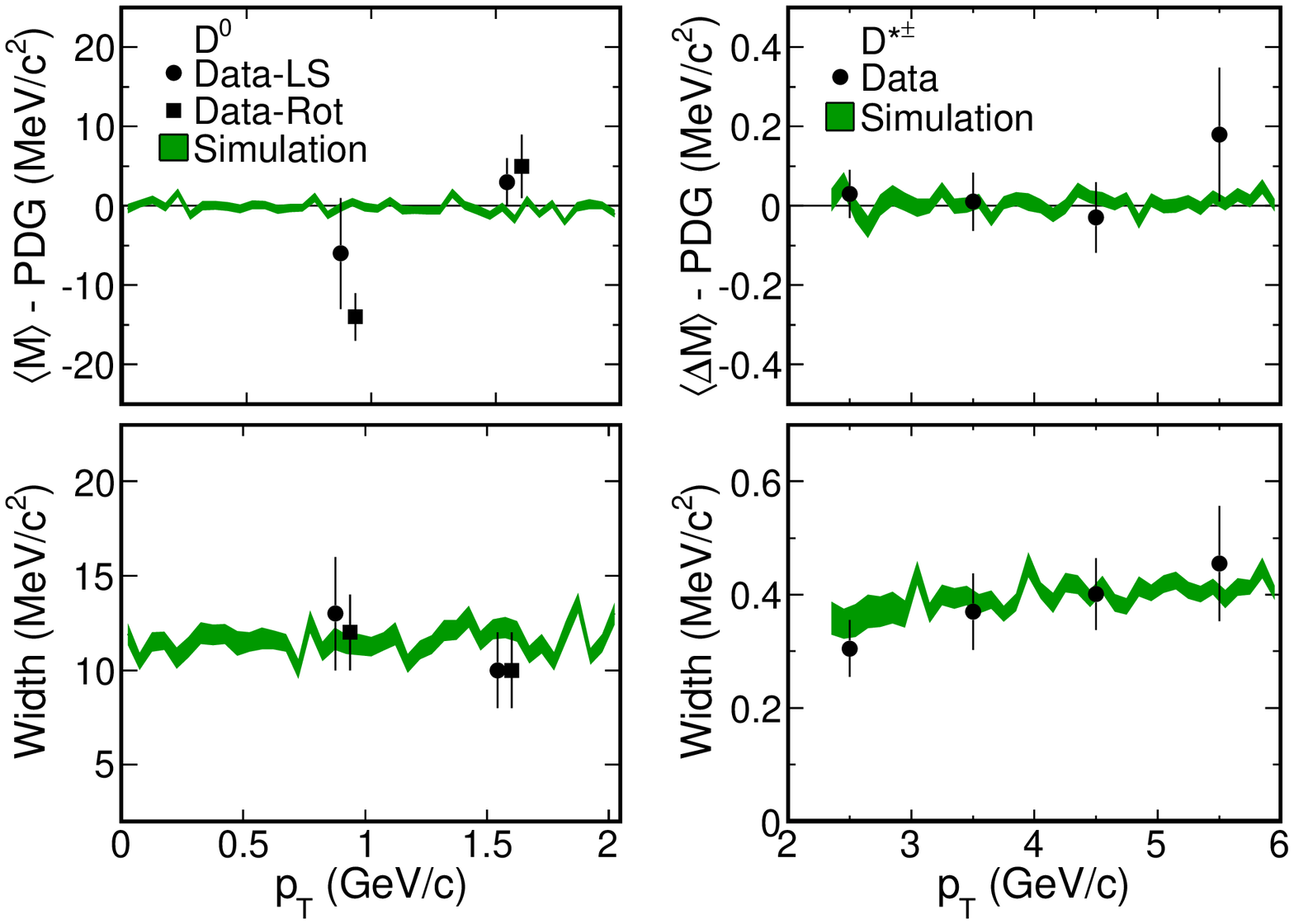}
\caption[]{The mean and width from Gaussian fit to data (symbols) compared with MC simulations (bands) for \Dzero and \Dstar are shown in left and right panels, respectively.}
\label{meanwidth}
\efg

\subsection{\Dstar Reconstruction}
$D^{*\pm}$ mesons were reconstructed via the decay sequence $D^{*+}\rightarrow D^0\pi^+$ ($BR$ = 67.7\%), $D^0\rightarrow K^-\pi^+$ and its charge conjugate. We followed the same analysis technique as described in Ref.~\cite{starDstar}. The daughter particles were still identified by $dE/dx$ in the TPC because a) most of the \Dstar decay daughter particles that fall inside the STAR acceptance with higher momenta are located in the region where the TOF PID improvement is very limited; and b) the signal suffers significant losses due to incomplete TOF acceptance in 2009. Compared to the cuts used in Ref.~\cite{starDstar}, the $p_T$ threshold cut for the $\pi^+$ (from \Dstar decays), denoted as $\pi^+_s$, was lowered to 0.15 GeV/$c$. The ratio, $r$, of transverse momenta from the $D^0$ and $\pi^+_s$ was required to be 7 $<r<$ 20. These two changes were implemented to improve the statistics near the lower bound in $p_T$. The remainder of the analysis cuts were the same as those used in Ref.~\cite{starDstar}.

\begin{figure}[floatfix] 
\centerline{\includegraphics[width=0.45\textwidth]{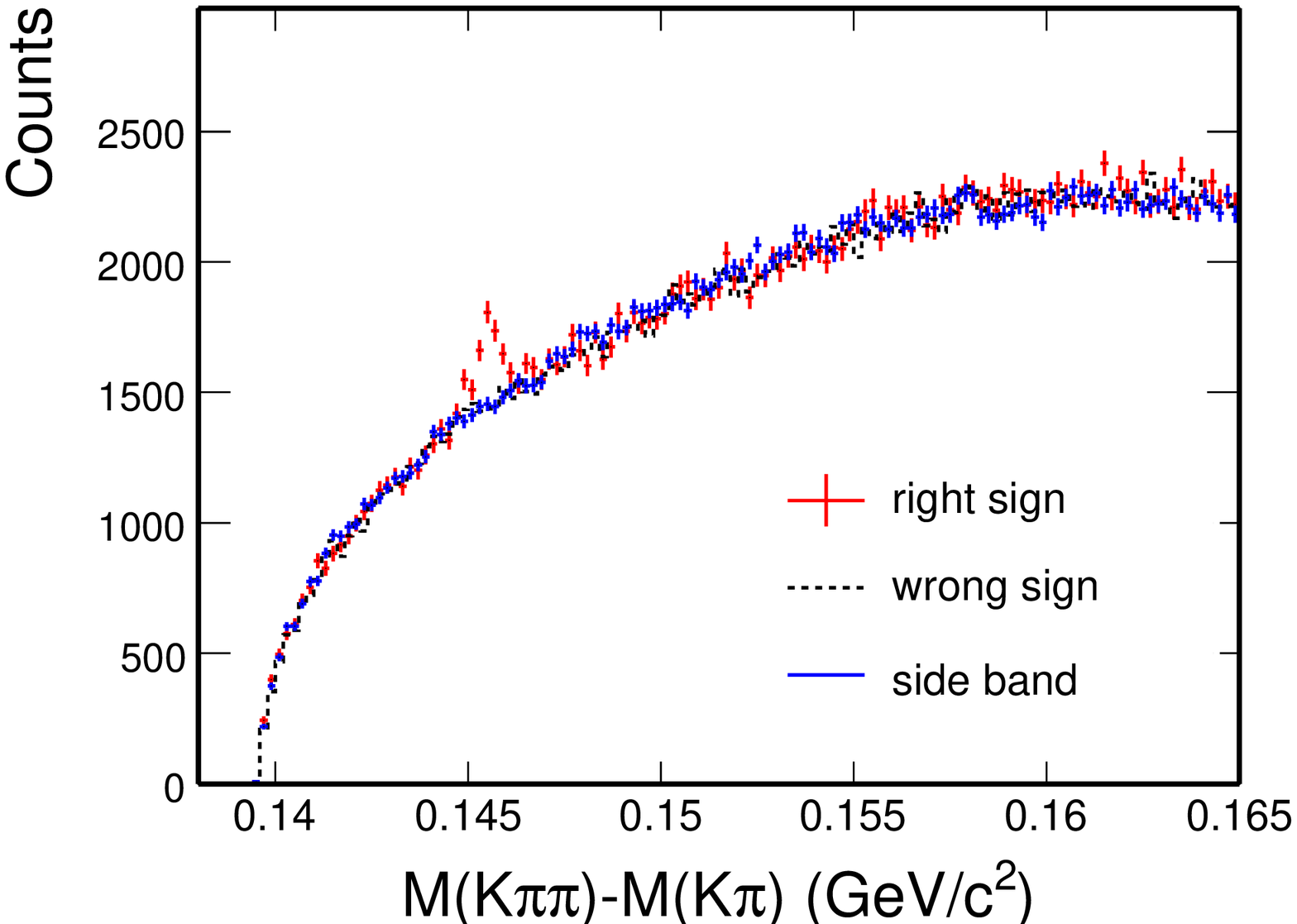}}
\centerline{\includegraphics[width=0.45\textwidth]{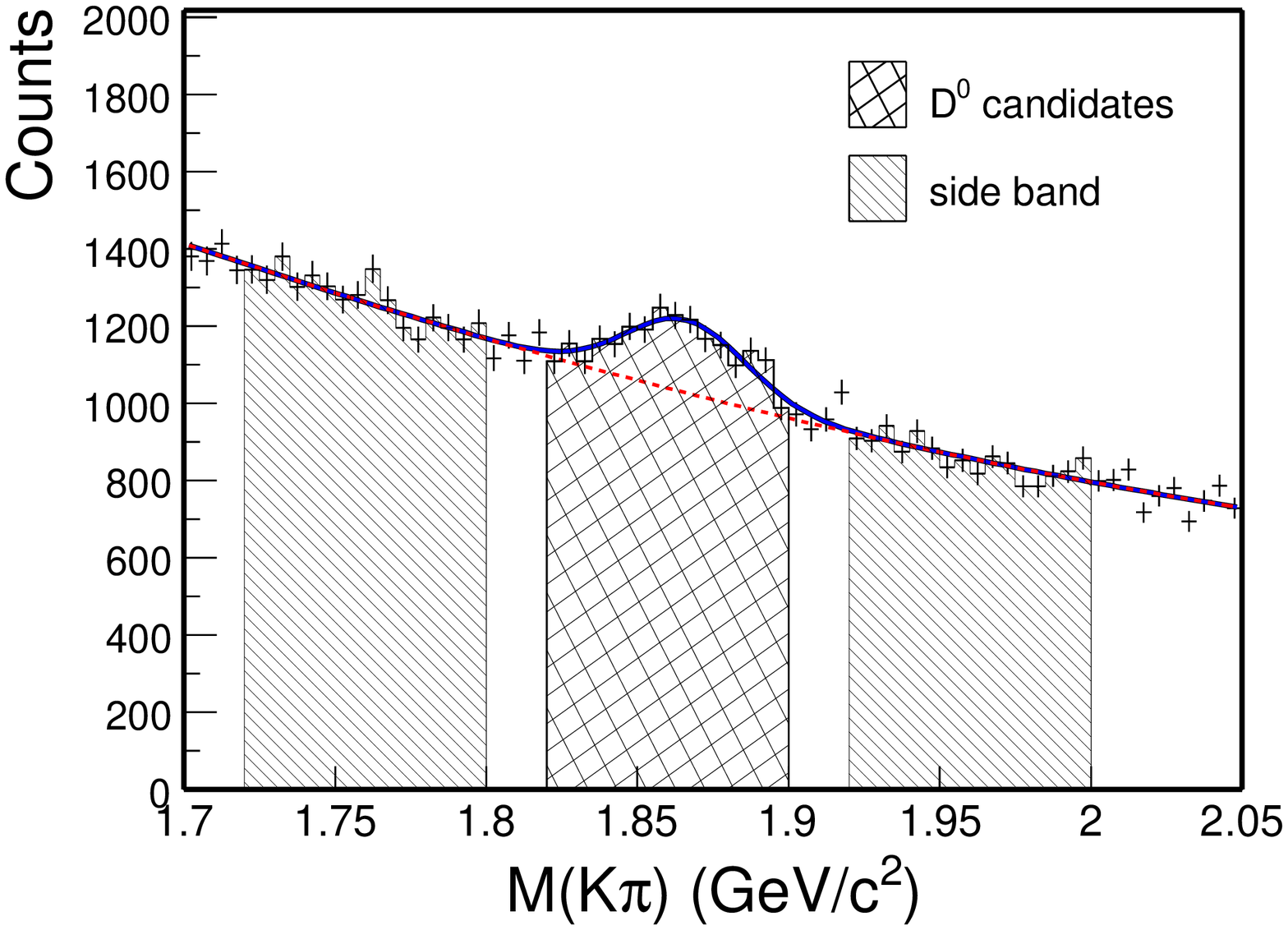}} 
\caption[]{(color online) Upper: Raw \Dstar candidate signal from the right-sign combinations in all \pp minimum-bias events. Histograms are combinatorial background distributions from ``wrong-sign'' and ``side-band'' methods. Lower: Raw $D^0$ candidates after requiring the \Dstar candidate cut (0.144 $<\Delta M<$ 0.147 GeV/$c^2$). } 
\label{fig6signal}
\end{figure}

The invariant mass difference $\Delta M = M(K\pi\pi)-M(K\pi)$ was
calculated in reconstructing the \Dstar signal to take advantage of the partial cancellation in the detector resolution in measured mass distributions. The $\Delta M$ distributions are shown in the upper panel of Fig.~\ref{fig6signal}. The
``right-sign'' combinations $K^\mp\pi^\pm\pi_s^\pm$ were used to
select the $D^{*\pm}$ candidates. Two independent methods --
``wrong-sign'' combinations $K^\pm\pi^\mp\pi_s^\pm$ and \Dzero
``side-band'' combinations -- were used for combinatorial background
reconstruction. The plot illustrates that both methods reproduce the
combinatorial background very well. The events displayed in this
figure are all minimum-bias events without event vertex selections, which
demonstrates the significance of \Dstar signal. The lower panel in
Fig.~\ref{fig6signal} shows the $K\pi$ invariant mass distribution
after requiring the \Dstar candidate cut (0.144 $<\Delta M<$ 0.147
GeV/$c^2$).  The cross-hatched area indicates $D^0$ candidate mass
selection in the $K\pi\pi$ right-sign and wrong-sign combination
reconstruction. The line-hatched area indicates the $D^0$ side-band
region (1.72 $< M(K\pi)/$(GeV/$c^{2}$) $<$ 1.80 or 1.92 $<
M(K\pi)/$(GeV/$c^{2}$) $<$ 2.00) used in side-band combinatorial
background reconstruction for \Dstar. The side-band combinatorial
background was used to obtain the raw \Dstar yields for better
statistics and also because side-band distributions do not suffer from the
double-counting issue due to particle misidentification. The difference between the
yields obtained from the side-band method and the ``wrong-sign'' method was included in the
systematic uncertainties. Details in determining the uncertainties on
the raw \Dstar yields including the double-counting effect will be
discussed in Sect. V A. The \Dstar raw yields are summarized in Table~\ref{Dstaryields}.

\begin{table}[htbp]
\caption[]{$D^{*}$ raw yields.} \label{Dstaryields} \vskip 0.1 in
\centering\begin{tabular}{|c|c|c|c|c|} \hline
$p_T$ range (GeV/$c$) & 2$-$3 & 3$-$4 & 4$-$5 & 5$-$6 \\ \hline
$p_T$ (GeV/$c$) & 2.45 & 3.44 & 4.45 & 5.45 \\ \hline
raw yields & 209 $\pm$ 58 & 98 $\pm$ 35 & 27 $\pm$ 11 & 12.3 $\pm$ 4.1 \\ 
\hline
\end{tabular}
\end{table}

To obtain the cross section, the event-selection criteria
described in the previous section were applied. The raw distributions
were further divided into $p_T$ slices to obtain the raw \Dstar yields
in each $p_T$ bin. Figure~\ref{fig7signalpT} shows the \Dstar
candidates and background distributions in different $p_T$ bins. The
bottom panel on each plot was generated by subtracting the
``side-band'' background from the ``right-sign'' candidates. The mean
and width from Gaussian fits are compared with MC simulation in the
right panel of Fig.~\ref{meanwidth}, and it shows the obtained \Dstar
peak positions and widths agree with the MC simulation well. From
this analysis, the total signal consisted of 364 $\pm$ 68 counts, and
the raw yield ratio of $D^{*-}/D^{*+}$ is 0.93 $\pm$ 0.37.


\bfg 
\centering \bmn[b]{0.25\textwidth} \centering
\includegraphics[width=1.0\textwidth]{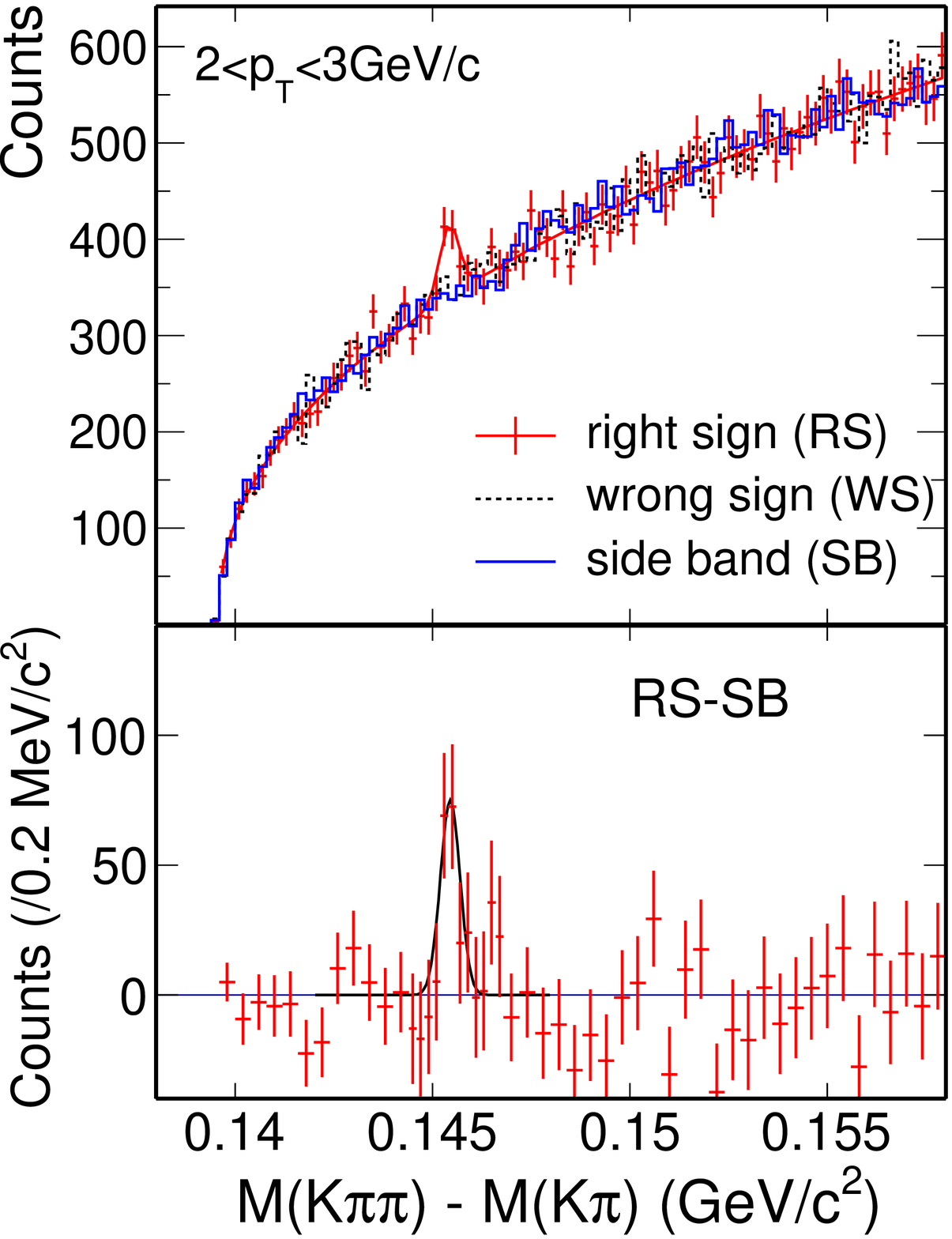}
\emn
\bmn[b]{0.25\textwidth} \centering
\includegraphics[width=1.0\textwidth]{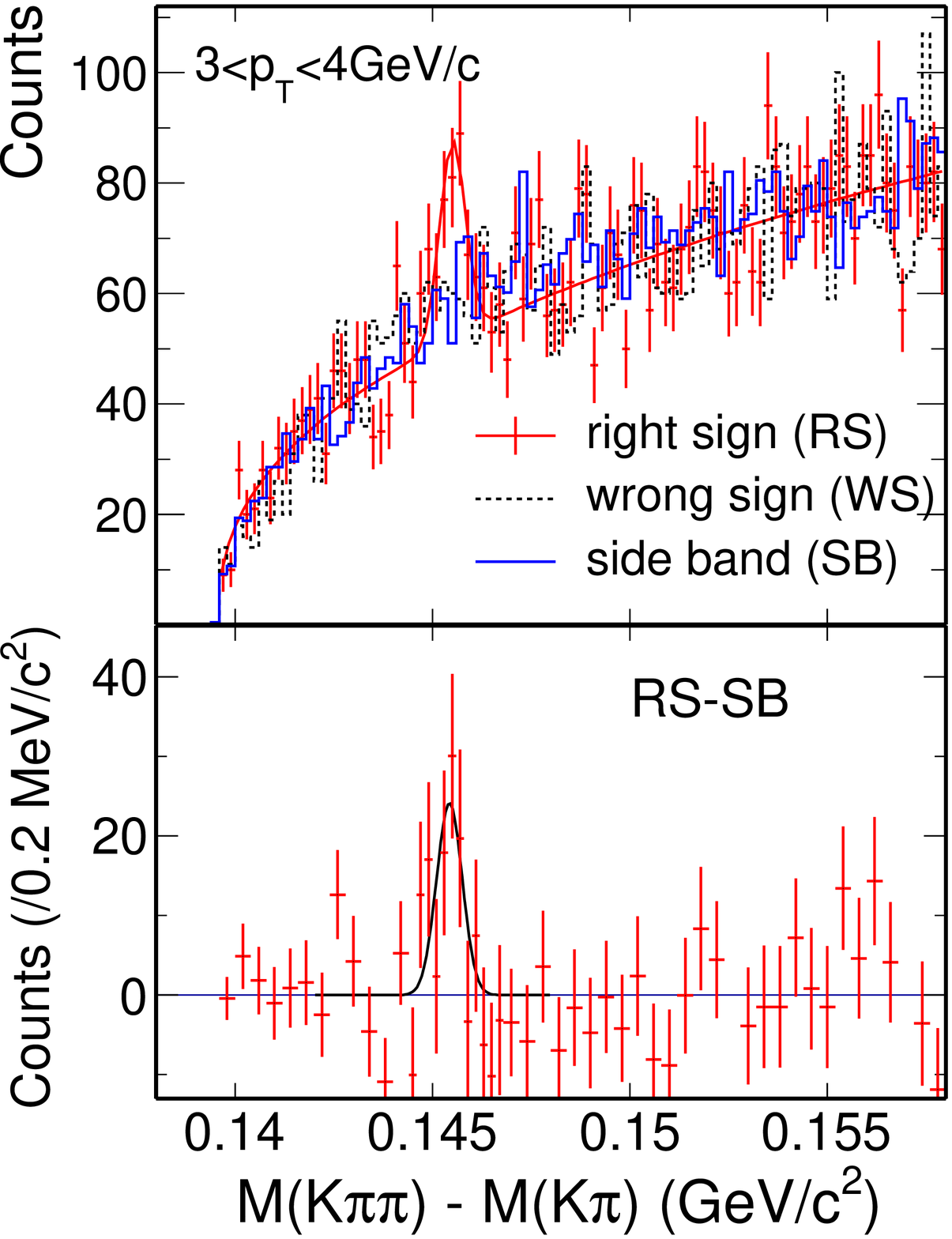}
\emn \\
\centering \bmn[b]{0.25\textwidth} \centering
\includegraphics[width=1.0\textwidth]{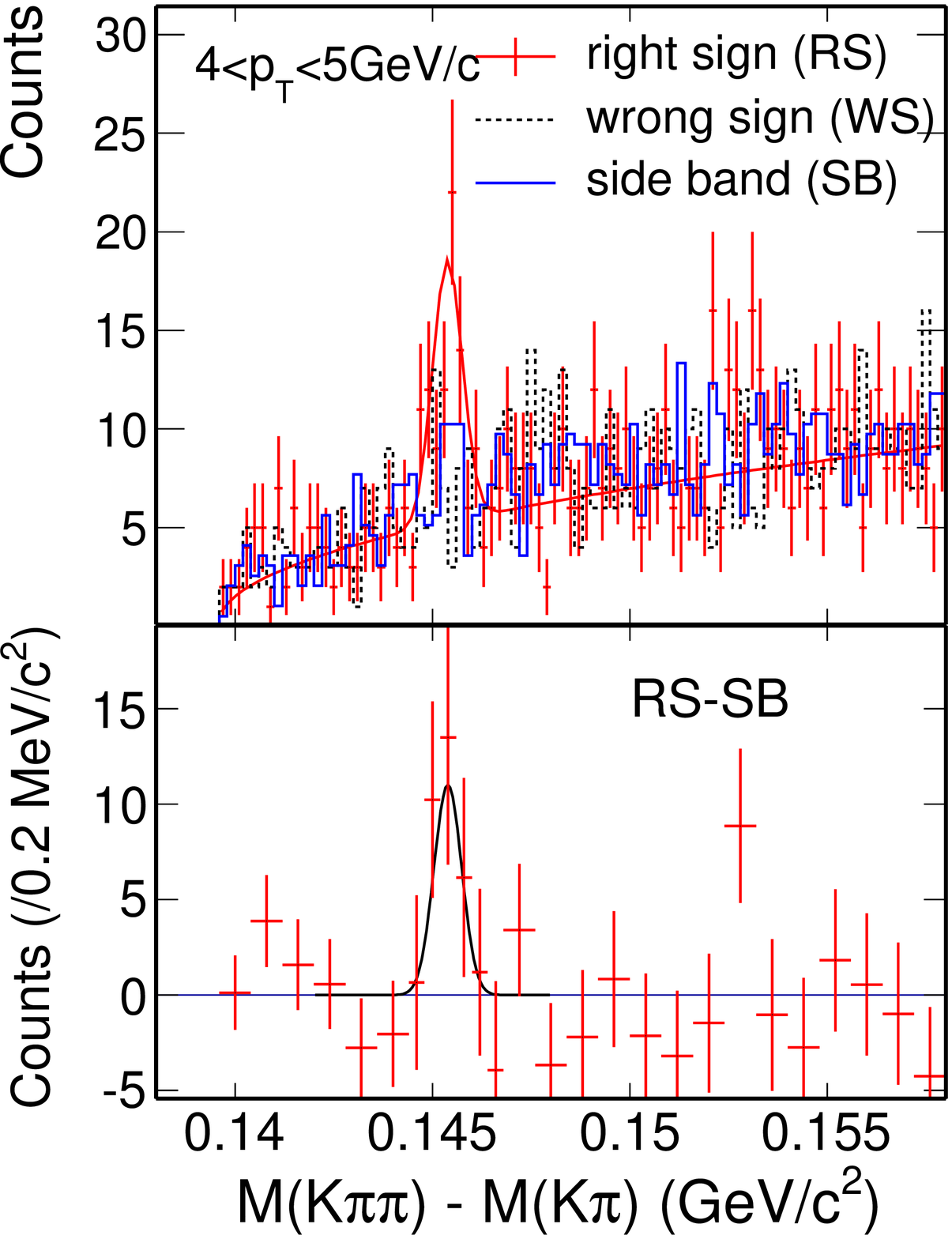}
\emn
\bmn[b]{0.25\textwidth} \centering
\includegraphics[width=1.0\textwidth]{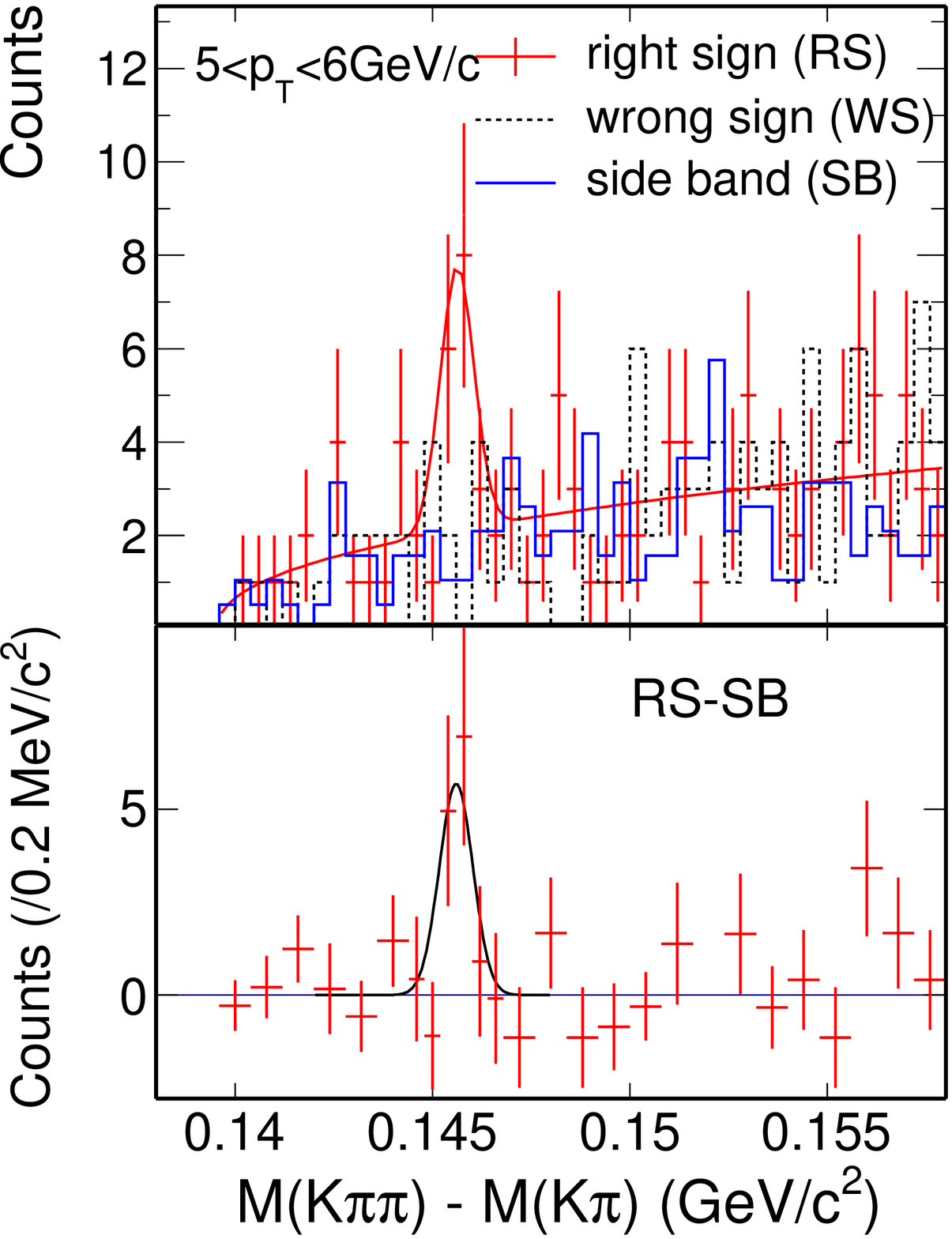}
\emn
\caption{(color online) Raw \Dstar signals in different $p_T$ bins. In each plot, the bottom panel distribution is generated by subtracting the side-band background from the right-sign distribution. Variable binning is used in the bottom panel for better illustration.}
\label{fig7signalpT} \efg

\section{Efficiency and Trigger/Vertex Bias Correction}

The final charmed-hadron cross section in \pp collisions is calculated as follows:

\begin{equation}
E\frac{d^{3}\sigma}{dp^3} = \frac{1}{2\pi} \cdot \frac{1}{\epsilon_{\text{rec}}}  \cdot \frac{1}{BR} \cdot \frac{\Delta N_{D}}{p_T \Delta p_T \Delta y} \cdot \frac{\sigma_{\text{NSD}}}{N_{\text{MB}}} \cdot f_{\text{trg,vtx}}.
\end{equation}

where $\sigma_{\text{NSD}}$ is the total Non-Singly Diffractive (NSD) cross section, which is measured at STAR to be 30.0 $\pm$ 2.4 mb~\cite{NSDXsec}. $N_{\text{MB}}$ is the total number of minimum-bias events used for the analysis. $\Delta N_{D}$ is the raw charmed-hadron signal in each $p_T$ bin within a rapidity window $\Delta y$. $BR$ is the hadronic decay branching ratio for the channel of interest. There are two correction factors: $\epsilon_{\text{rec}}$, which is the reconstruction efficiency including geometric acceptance, track selection efficiency, PID efficiency, and analysis cut efficiency; and $f_{\text{trg,vtx}}(p_T)$, which is the correction factor to account for the bias between the minimum-bias sample used in this analysis and the total NSD sample. This bias is mainly caused by the VPD trigger and event vertex reconstruction, and it may have a dependence on the charmed-hadron $p_T$. In the following sections of the paper, the condition that requires the event to fire the VPD trigger and to have a good vertex will be referred to as the ``Analysis Condition''.

\subsection{Reconstruction Efficiency}

The reconstruction efficiency for charmed hadrons was obtained by embedding Monte Carlo (MC) simulated charmed-hadron tracks into the real minimum-bias events. The MC charmed-hadron tracks were processed through a full GEANT detector simulation~\cite{geant:321} with a representation of the 2009 STAR geometry. The raw detector-response signals were mixed together with those from the real data and processed through the full STAR offline reconstruction chain to obtain the detector response efficiency in a realistic environment. The input MC track multiplicity was constrained to have negligible effect on the final tracking efficiency due to increased occupancy in the TPC.

Figures~\ref{fig8effidzero} and~\ref{fig9effidstar} show the \Dzero and \Dstar reconstruction efficiency versus $p_T$ within $|y|<$1. In Fig.~\ref{fig8effidzero}, the solid squares denote the reconstruction efficiency for both daughters selected and identified by the TPC, while the solid circles denote the reconstruction efficiency with additional PID selection from the TOF detector for the kaon daughter. The combined TOF efficiency, including the acceptance, matching between TPC tracks and TOF hits, and PID selection efficiency, is around 45\% studied from the data in 2009.

\bfg \centering
\includegraphics[width=0.45\textwidth]{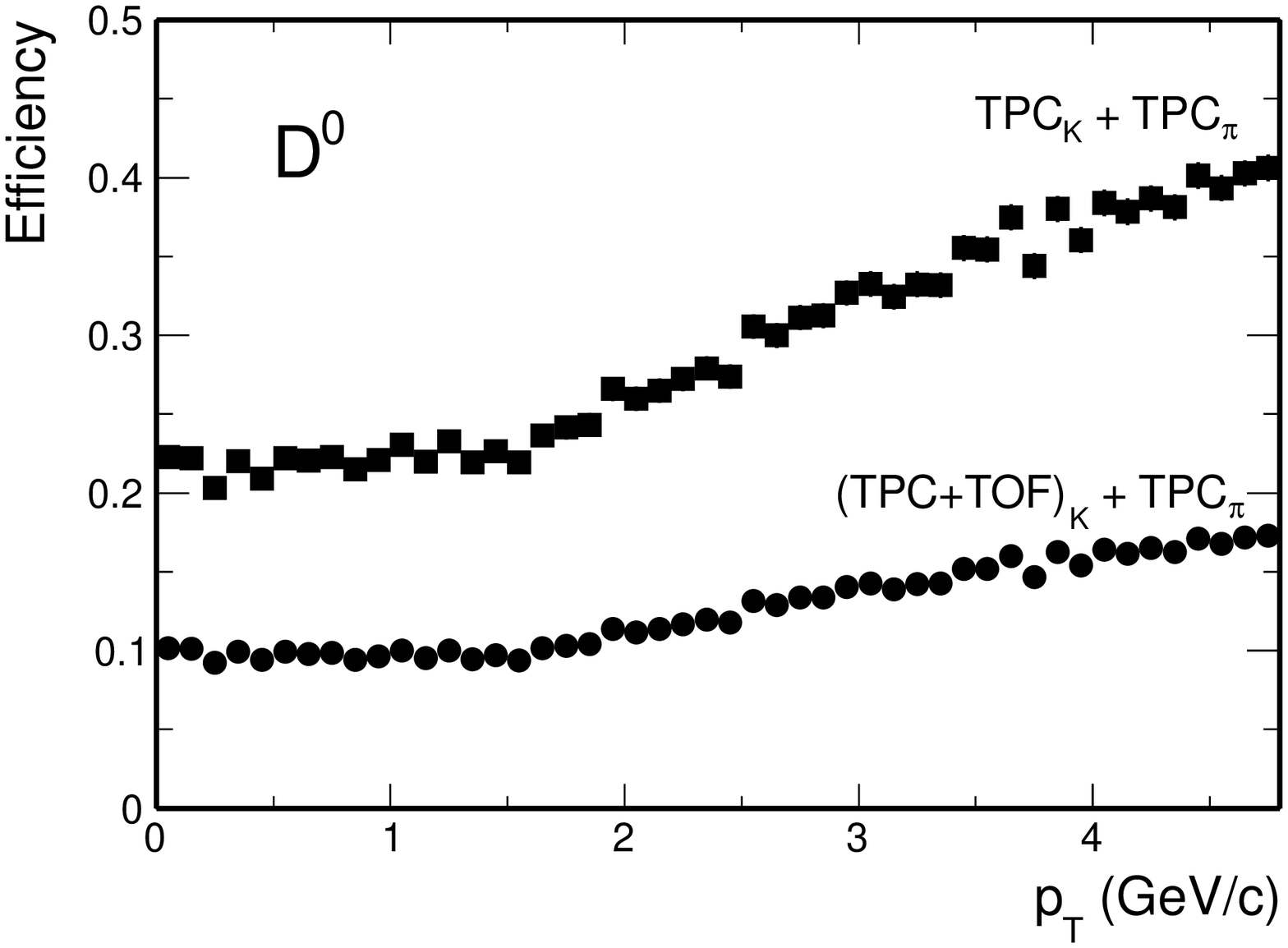}
\caption[]{Total \Dzero reconstruction efficiency versus $D^{0}$ $p_T$.}
\label{fig8effidzero}
\efg

\bfg \centering
\includegraphics[width=0.45\textwidth]{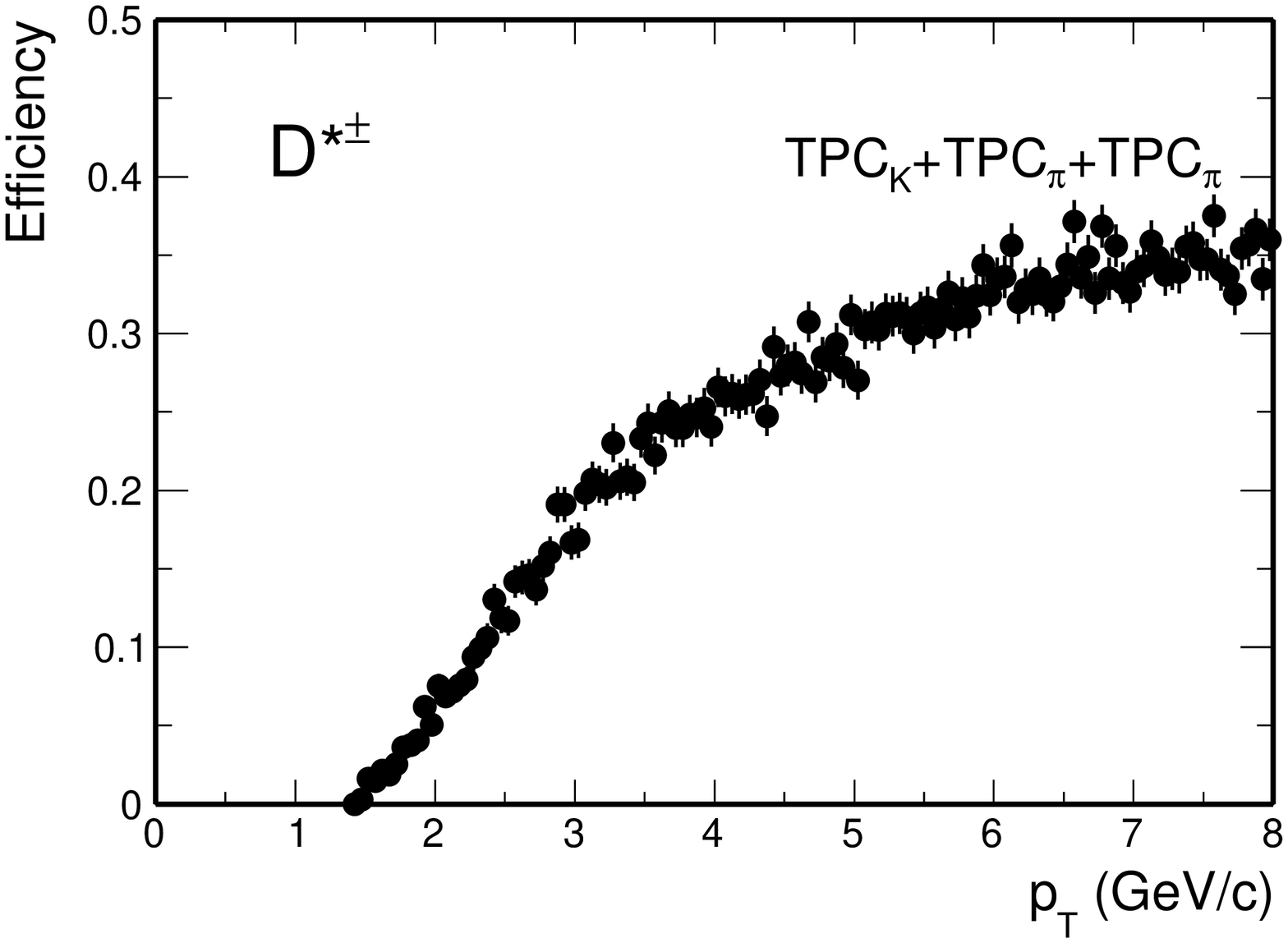}
\caption[]{Total \Dstar reconstruction efficiency versus $D^{*}$ $p_T$.}
\label{fig9effidstar}
\efg

\subsection{Trigger and Vertex Bias Corrections}
The trigger and vertex bias corrections were studied by simulating
PYTHIA events~\cite{pythia} processed through the full GEANT detector
response and offline reconstruction. The PYTHIA generator versions
6.205 and 6.416 were both used in this study. We chose the PYTHIA version
6.205 with minimum-bias processes selected and with the CDF TuneA
settings~\cite{CDF:TuneA} to give the centroid value of the
correction factor because it gives better description for the particle production in the forward
rapidities than the 6.416 version~\cite{transversity}. The differences between the two
versions as well as different parameter settings have been included to
estimate the systematic uncertainty of the trigger and vertex bias correction factor.

To validate the PYTHIA generator in simulating particle production in the forward region for the VPD trigger study, we first compared the VPD trigger efficiencies (from the BBC triggered minimum-bias sample) from MC simulation and real data. The BBC trigger has been well studied and was used to calculate the \pp NSD cross section~\cite{STARnpepp}. Figure~\ref{fig10simucomp} shows the comparison of the VPD trigger efficiency, with the requirement that there is a BBC trigger and a good vertex. The efficiency is studied as a function of the charged hadron $p_T$. The real data used are BBC triggered minimum-bias events taken in 2009 during a very low luminosity run, which minimizes TPC pileup tracks. Figure~\ref{fig10simucomp} shows that the efficiency goes down with increasing \pt of mid-rapidity particles indicating an anti-correlation between mid-rapidity particle production and forward VPD triggering. Most importantly, within the momentum range under study, the PYTHIA MC simulation agrees well with the data. This agreement provides confidence in using PYTHIA simulations to evaluate this correction. 

\bfg \centering
\includegraphics[width=0.45\textwidth]{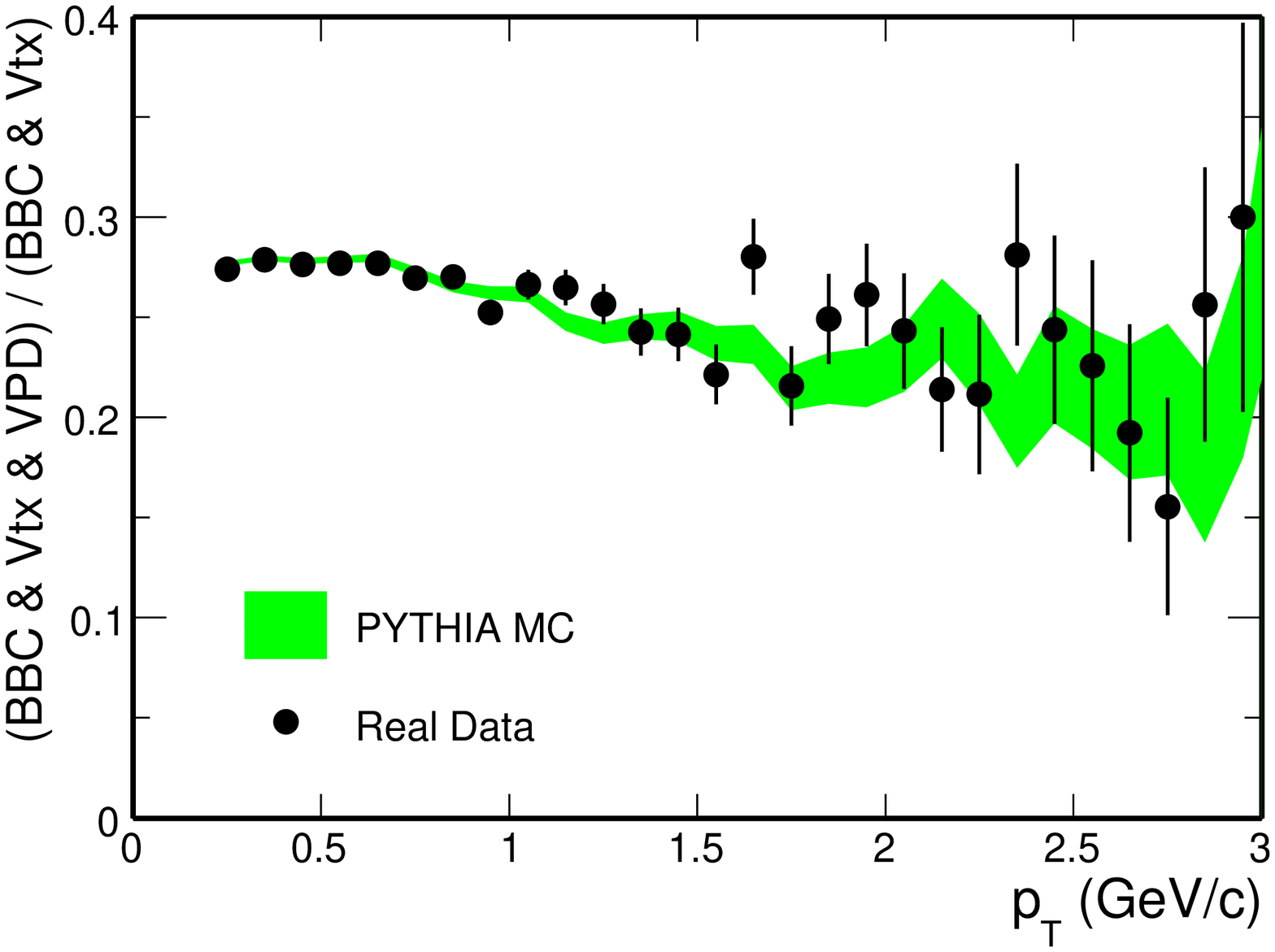}
\caption[]{VPD trigger efficiency comparison between Data and Monte Carlo versus charged particle $p_T$ in BBC MB conditions.}
\label{fig10simucomp}
\efg

The correction factor $f_{\text{trg,vtx}}$ can be related to the ratio $(N_D/N_{\text{mb}})$ for the pure minimum-bias condition and the ``Analysis Condition'', $i.e.$
\begin{equation}
f_{\text{trg,vtx}}(p_T) \equiv \frac{N_D(p_T)/N_{\text{mb}}}{N_D^{\text{trg,vtx}}(p_T)/N_{\text{mb}}^{\text{trg,vtx}}}.
\end{equation}

Two simulation samples were generated to obtain the correction factor. One sample consisted of PYTHIA-simulated \pp events and was used to obtain the fraction of minimum-bias events that satisfy the ``Analysis Condition'' : $N_{\text{mb}}^{\text{trg,vtx}}/N_{\text{mb}}$. This fraction was found to be 12.7\% from this PYTHIA simulation. The other simulation sample was generated using the same PYTHIA settings, but only events with at least one charmed hadron were saved to enhance the statistics. This sample was used to obtain the fraction of charmed-hadron signals that satisfy the ``Analysis Condition'' - $N_D^{\text{trg,vtx}}/N_D$. We also studied this fraction as a function of charmed-hadron $p_T$. Figure~\ref{fig11trgvtxeffcharm} shows the calculated efficiencies for \Dstar from different event-selection criteria. The BBC coincidence study provides a baseline for this simulation, which demonstrates consistency with previous STAR results~\cite{NSDXsec}. As expected, the vertex finding efficiency increases with increasing $p_T$. The VPD trigger efficiency shows an anti-correlation with increasing \Dstar $p_T$, similar to that observed with increasing charged-hadron $p_T$. The final efficiency (with requirements for both vertexing and VPD triggering) is almost flat versus $p_T$, leveling off at $\sim$19\%. The simulation for \Dzero hadrons shows very similar results. Figure~\ref{fig12trgvtxcorr} shows the correction factor, $f_{\text{trg,vtx}}$, for cross section calculations for \Dzero and $D^{*}$.

\bfg \centering
\includegraphics[width=0.45\textwidth]{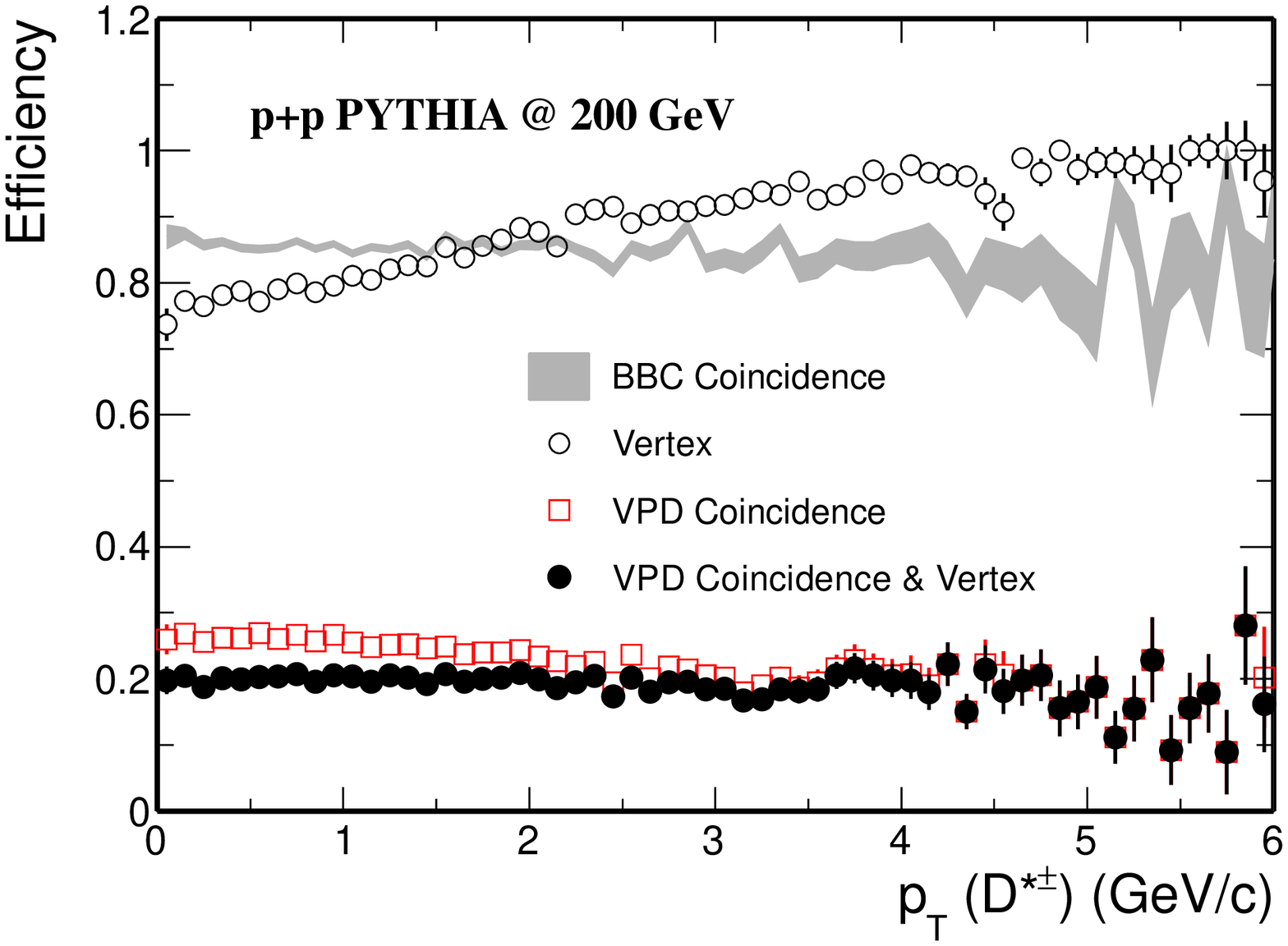}
\caption[]{\Dstar efficiency versus \Dstar \pt with different event selection criteria.}
\label{fig11trgvtxeffcharm}
\efg

\bfg \centering
\includegraphics[width=0.45\textwidth]{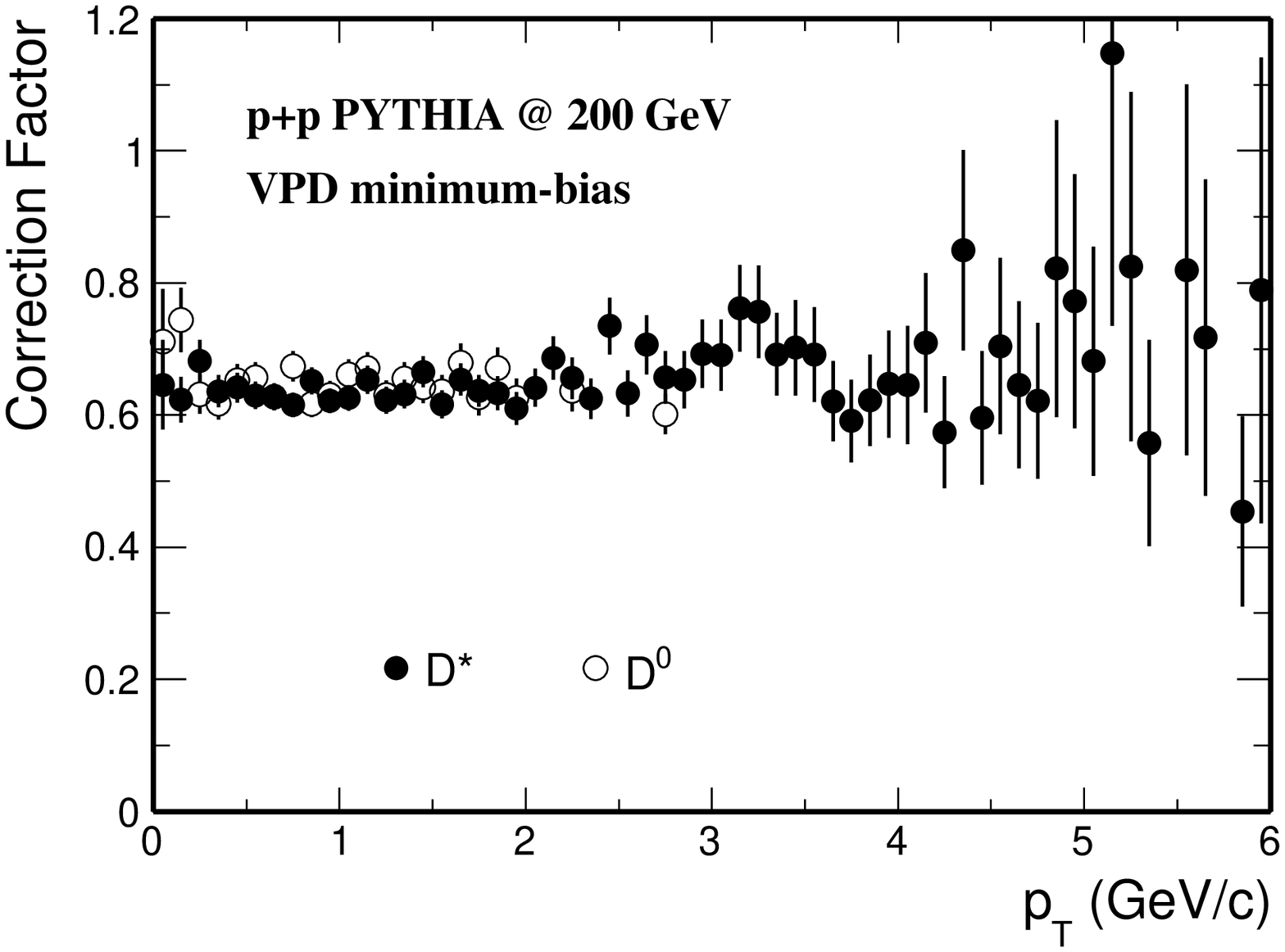}
\caption[]{The correction factor $f_{\text{trg,vtx}}$ versus charmed-hadron \pt for cross section calculations for \Dzero and \Dstar.}
\label{fig12trgvtxcorr}
\efg

\section{Systematic Uncertainties}

Sources that contribute to the systematic uncertainties in the final
$D$-meson cross sections include: \textit{a}) uncertainty in determining the raw
$D$-meson yields; \textit{b}) uncertainty in determining the
reconstruction efficiency; \textit{c}) uncertainty of the total NSD
cross section and \textit{d}) uncertainty in determining the
trigger/vertex correction factor. Uncertainties due to particle
identifications will enter in both \textit{a}) and \textit{b}) which
will be discussed in the following subsections. 
We consider \textit{a}) as point-by-point uncorrelated systematic uncertainties.
Although \textit{b}) is correlated in $p_T$, it is not simply a normalization uncertainty, and the exact correlation in $p_T$ is not known.  Therefore we include \textit{b}) in the point-by-point uncorrelated systematic uncertainties.
Finally, \textit{c}) and \textit{d}) are overall normalization uncertainties.

\subsection{Uncertainty in Raw Yields}
Different choices on background reconstruction methods, function fits
and mass binning were used to evaluate the
systematic uncertainty in the raw $D$-meson yields. In the \Dzero
analysis, the difference
between the yields extracted from Rot and LS methods is 15.6$-$18.9\%. Fitting the
\Dzero peak with fixed parameters from simulation estimates lower
yields of 28.2\% and 6.1\% for the two \Dzero \pt bins. The systematic uncertainties from different mass binning
and different fit regions are estimated to be $\sim$5$-$7\%. 
The systematic uncertainties in determining the raw
\Dstar yields include contributions from the difference obtained between the ``side-band''
and the ``wrong-sign'' methods,  and the difference between bin counting and Gaussian
fitting methods, varying $\sim$6$-$11\% in the \pt range 2$-$6 GeV/$c$. The
choice of mass binning and fitting range had a negligible effect on the extracted yields.

In \Dzero meson reconstruction, if the kaon (pion) daughter is misidentified as a pion (kaon) then two daughters from a real \Dzero decay will show up as additional $\overline{D^0}$ combinations with a wider mass distribution due to wrong mass assignments. Thus one \Dzero signal will be counted twice; once as a \Dzero and again as a $\overline{D^0}$. A Monte Carlo simulation was used to evaluate the fraction of such double counting occurrences in the \Dzero reconstruction. Based on realistic $dE/dx$ and TOF PID resolutions extracted from real data, the probability that kaons (pions) can be misidentified as pions (kaons) at a given $p_T$, using these PID selections, was obtained. Assuming a $D^0$ candidate, this procedure provides an estimate of the probability that both daughters are misidentified and then reconstructed as a $\overline{D^0}$. In Fig.~\ref{fig13doubleCountingFrac}, the open and closed circles show the double-counting fraction, relative to the total real signal, for two different PID selections: a) both daughters are identified by TPC $dE/dx$; b) the kaon daughters are identified by the TOF, while pions are identified by the TPC. The sharp increase at very low $p_T$ (identifying both daughters using $dE/dx$) is due to the case where a \Dzero decays almost at rest ($p_T\sim$0), and the two daughters are produced in the momentum region where the kaon and pion $dE/dx$ bands cross, therefore maximizing the misidentification probability. The plot shows that when the kaon daughter is identified by the TOF, the double-counting fraction is negligible in our \Dzero $p_T$ coverage region (0.6$-$2.0 GeV/$c$).

\bfg \centering
\includegraphics[width=0.45\textwidth]{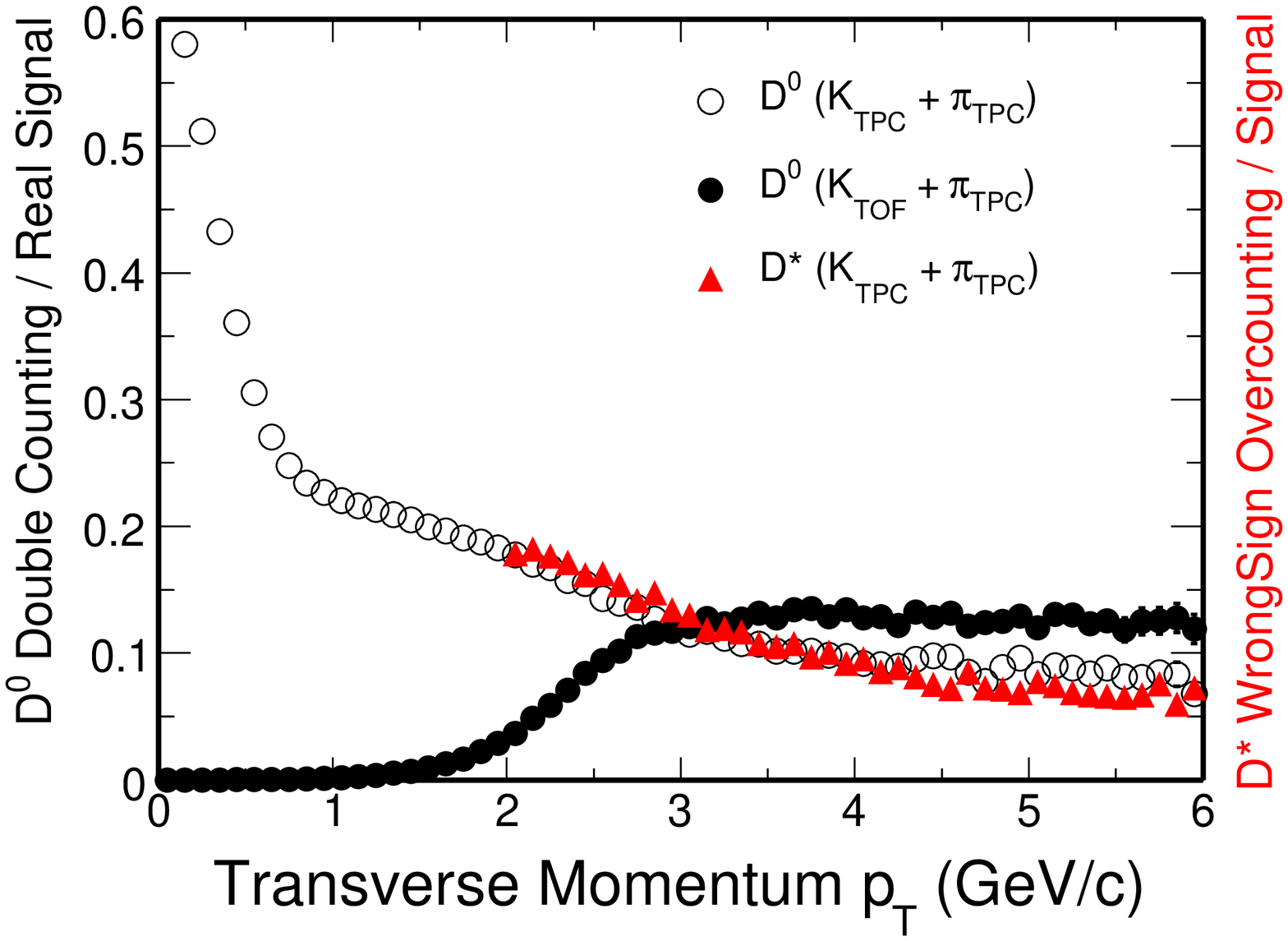}
\caption[]{\Dzero double-counting fraction due to particle misidentification in two PID selections and \Dstar wrong-sign over counting fraction versus $D$-meson $p_T$.}
\label{fig13doubleCountingFrac}
\efg

Double counting the \Dzero may also impact reconstruction of
$D^{*}$. However, the impact is different because of a charge sign
requirement on the soft pions. If both daughters from a \Dzero are
misidentified (\Dzero is reconstructed as $\overline{D^0}$), then the
combination from the same signal will become $K^+\pi^-\pi^+$. It will
not contribute to the right-sign distributions, but instead, will
enter into the ``wrong-sign'' (background) distributions if the  mass
also falls into the \Dzero ($\overline{D^0}$) mass selection
window. Thus the double counting in ``wrong-sign'' background will
contribute to an undercounting in the total signal if the wrong-sign
background is subtracted from the right-sign distribution. Since the
right-sign combination was also required, the misidentification does
not affect the side-band background distributions. In the real
analysis, the side-band background subtraction was used to extract the
raw signal, but also the difference between side-band and wrong-sign
methods was used for systematic uncertainty estimation. Since the
wrong-sign distribution can be overestimated due to particle
misidentification, the systematic error from the difference between
the two methods would be overestimated. This was avoided with better
understanding of the wrong-sign overcounting. The red triangles in
Fig.~\ref{fig13doubleCountingFrac} denote the over counting fraction
in the \Dstar wrong-sign background to real signals. It is very close
to the \Dzero double-counting fraction, since they are from the same
source. The slight difference comes from the additional \Dzero
candidate selection cuts used in the \Dstar reconstruction. This fraction was used to compensate for the difference between the two background methods and as a way to improve the assessment of the systematic uncertainties in the extraction of the raw \Dstar yields.

\subsection{Uncertainty in Reconstruction Efficiency}

The systematic uncertainties of the reconstruction efficiencies were
obtained following similar methods used in other particle cross
section measurements by changing the daughter track selection criteria
and comparing the difference between the data and the MC. In this
analysis, it was studied by changing the minimum number of fit points (nFitPts) in the TPC from 15 to 25
and the DCA to the collision vertex from 2 cm to 1 cm. The uncertainty was then
quantified by the difference in the remaining fractions after cut changes between
the data and the MC. For each cut
change, the uncertainties were calculated for each decay daughter and
added together linearly to obtain the total for \Dzero and
$D^{*}$. The systematic uncertainties on the PID
cut efficiencies (from both $dE/dx$ and TOF) were estimated to be
$<$1\% and neglected in the total uncertainty. Then the uncertainties from the cut changes on nFitPts and DCA
were added in quadrature to obtain the total systematic uncertainty on the reconstruction efficiency.

The point-by-point systematic errors including uncertainties in raw
yields and reconstruction efficiency for the \Dzero and \Dstar cross
sections in each $p_T$ bin are summarized in Table~\ref{dstarsystable}.

\begin{table}[htbp]
\caption[]{$D^{0}$ (0.6$-$2 GeV/$c$) and $D^{*}$ (2$-$6 GeV/$c$) point-by-point systematic errors (\%)} \label{dstarsystable} \vskip 0.1 in
\centering\begin{tabular}{|c|c|c|c|c|c|c|} \hline
$p_T$ (GeV/$c$) &0.6$-$1.2 & 1.2$-$2 & 2$-$3 & 3$-$4 & 4$-$5 & 5$-$6 \\ \hline
raw yields & +18.9 & +15.6 & 9.4 & 6.5 & 11.0 & 6.6 \\ 
 & -33.9 & -16.8 & & & & \\ \hline
nFitPts 15$\rightarrow$25 & 3.8 & 3.2 & 7.2 & 4.7 & 5.9 & 4.7 \\ \hline
DCA 2$\rightarrow$1 (cm) & 6.6 & 7.1 & 13.6 & 12.7 & 11.6 & 10.7 \\ \hline
quadratic sum & +20.8 & +17.8 & 18.1 & 15.1 & 17.1 & 13.5 \\
 & -34.8 & -18.5 & & & & \\
\hline
\end{tabular}
\end{table}

\subsection{Overal Normalization Uncertainty}
The overall normalization uncertainty for the total NSD cross
section has been studied before and reported in previous STAR
publication~\cite{NSDXsec}. It was estimated to be 8.1\%, including
the uncertainty from measuring the absolute BBC cross section and that
of BBC triggering efficiency. The uncertainty from the trigger/vertex
bias correction factor amounts to 5.2\% by varying different
PYTHIA versions (6.205 vs. 6.416) and different parameter settings in
the simulation. We also considered the impact from pileup
TPC tracks as an additional systematic source on the correction factor
, and the uncertainty was estimated to be 4.0\% by comparing the result
with a conservative luminosity level for this data set to that from
pure PYTHIA simulation without pileup. 

These uncertainties were added in quadrature, which gives 10.4\% overall
normalization uncertainty for the $D$-meson cross sections.

\section{Result and discussion}
\begin{figure}[h]
\centerline{\includegraphics[width=0.45\textwidth]{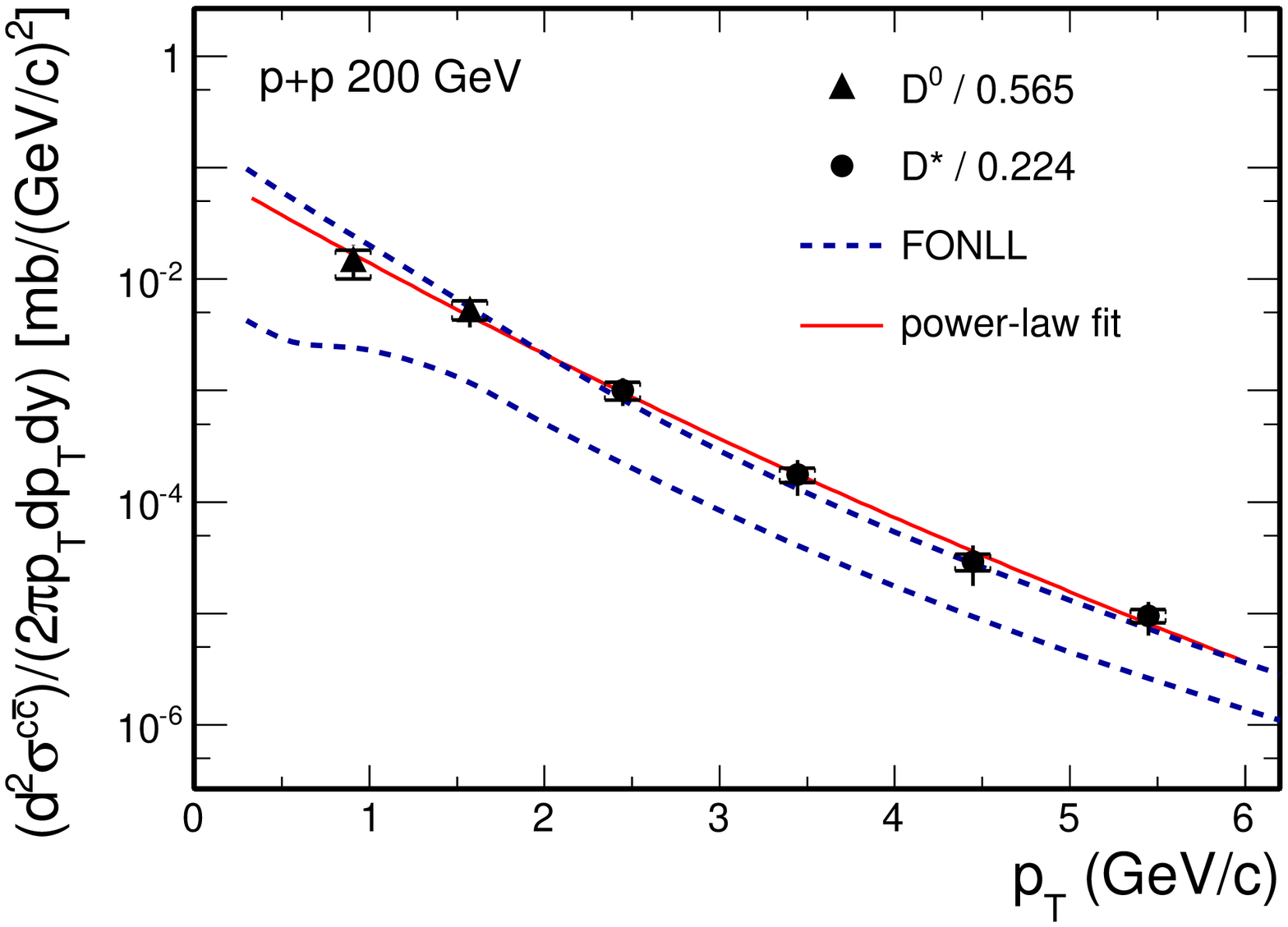}}
\caption[]{(color online) $c\bar{c}$ production cross section as inferred from \Dzero and \Dstar production in \pp collisions at \srt = 200 GeV compared with FONLL calculations. The \Dzero and \Dstar data points were divided by the charm quark fragmentation ratios 0.565 ($c\rightarrow D^0$) and 0.224 ($c\rightarrow D^{*+}$) ~\cite{pdg}, respectively, to convert to the $c\bar{c}$ production cross section.} 
\label{fig14spectra}
\end{figure}

After the reconstruction efficiency and trigger/vertex bias correction
factor were applied, the differential production cross sections for
\Dzero and \Dstar in \pp collisions at \srt = 200 GeV were extracted,
as shown in Fig.~\ref{fig14spectra}. The vertical bars on the data
points indicate the statistical uncertainties, while the brackets
indicate the bin-to-bin systematic uncertainties described in previous
section. The \Dzero and \Dstar cross sections were divided by the
charm quark fragmentation ratios 0.565 $\pm$ 0.032 ($c\rightarrow
D^0$) and 0.224 $\pm$ 0.028 ($c\rightarrow D^{*+}$), respectively, to
convert to the $c\bar{c}$ production cross section. The charm quark
fragmentation ratios are measured from CLEO and BELLE experiments near
the $\Upsilon$ resonance~\cite{pdg}. The uncertainties of the
fragmentation ratios are taken into account as systematic errors in
calculating the $c\bar{c}$ production cross section. A power-law fit to the data points was performed with the following function~\cite{dAuCharm}:

\begin{equation}
E\frac{d^3\sigma}{dp^3} = \frac{d\sigma}{dy}\frac{2(n-1)(n-2)}{\pi(n-3)^2\langle p_T\rangle^2}(1+\frac{p_T}{\langle p_T\rangle(n-3)/2})^{-n}.
\end{equation} and shown as the solid red line in the figure. The fit quality with the power-law function, measured as $\chi^2/{\rm ndf}$, is 0.9/3 with statistical errors and 3.7/3 with point-by-point systematic errors, respectively. The latter was used to extract the systematic uncertainty on the $p_T$ integrated cross section from point-by-point systematic sources. The obtained $c\bar{c}$ production cross section at mid-rapidity is,

\begin{equation}
\frac{d\sigma}{dy}|_{y=0}^{c\bar{c}} = 170 \pm 45~(\text{stat.})~ ^{+38}_{-59}~(\text{sys.})~ \mu\text{b}.
\end{equation}

The term with sys. includes the uncertainty arising from the bin-to-bin systematic uncertainties and from the extrapolation to the low-$p_T$ region, which is not measured. The FONLL upper limit and PYTHIA+tune fits are used for the low-\pt extrapolation, which gives +6.2\% and -16.4\% uncertainties, respectively. 
At mid-rapidity, about 67\% of the D meson yield falls in the measured \pt region. The mean transverse momentum of charmed mesons is found to be $1.06 \pm 0.14 ~(\text{stat.}) \pm 0.09 ~(\text{sys.})$~GeV/$c$. The charm-pair cross section at mid-rapidity from this measurement is consistent with STAR' s previous measurement in $d$+Au collisions~\cite{dAuCharm} at 1.7$\sigma$ ($\sigma$ is the averaged total uncertainty between two results), providing negligible nuclear effects in $d$+Au collisions.

Also shown in Fig.~\ref{fig14spectra} are the upper and lower edges (blue dashed lines) of a FONLL pQCD calculation taken from Ref.~\cite{FONLL}. Our results are consistent with the upper limit of the FONLL pQCD calculation in a wide $p_T$ region. It is observed that the charmed-hadron cross sections measured by CDF~\cite{CDFcharm} and ALICE~\cite{ALICEcharm} at energies up to 7 TeV are also close to the upper limits of FONLL pQCD calculations. This may help set constraints on the parameters used in the FONLL calculations, e.g. on the choice of renormalization or factorization scales, which are the main parameters varied to obtain the upper and lower limits on these calculations. However one should note the valid $p_T$ region of FONLL calculations when applying such a analysis since FONLL calculations are supposed to work when $p_T \gg m_c$.


The charm cross section at mid-rapidity was extrapolated to full phase space using the same extrapolation factor, 4.7 $\pm$ 0.7, as in a previous publication ~\cite{dAuCharm}, and the extracted charm total cross section at \srt = 200\,GeV is

\begin{equation}
\sigma_{c\bar{c}} = 797 \pm 210 ~(\text{stat.}) ^{+208}_{-295} ~(\text{sys.}) ~\mu\text{b}.
\end{equation}

\begin{figure}[h]
\centerline{\includegraphics[width=0.45\textwidth]{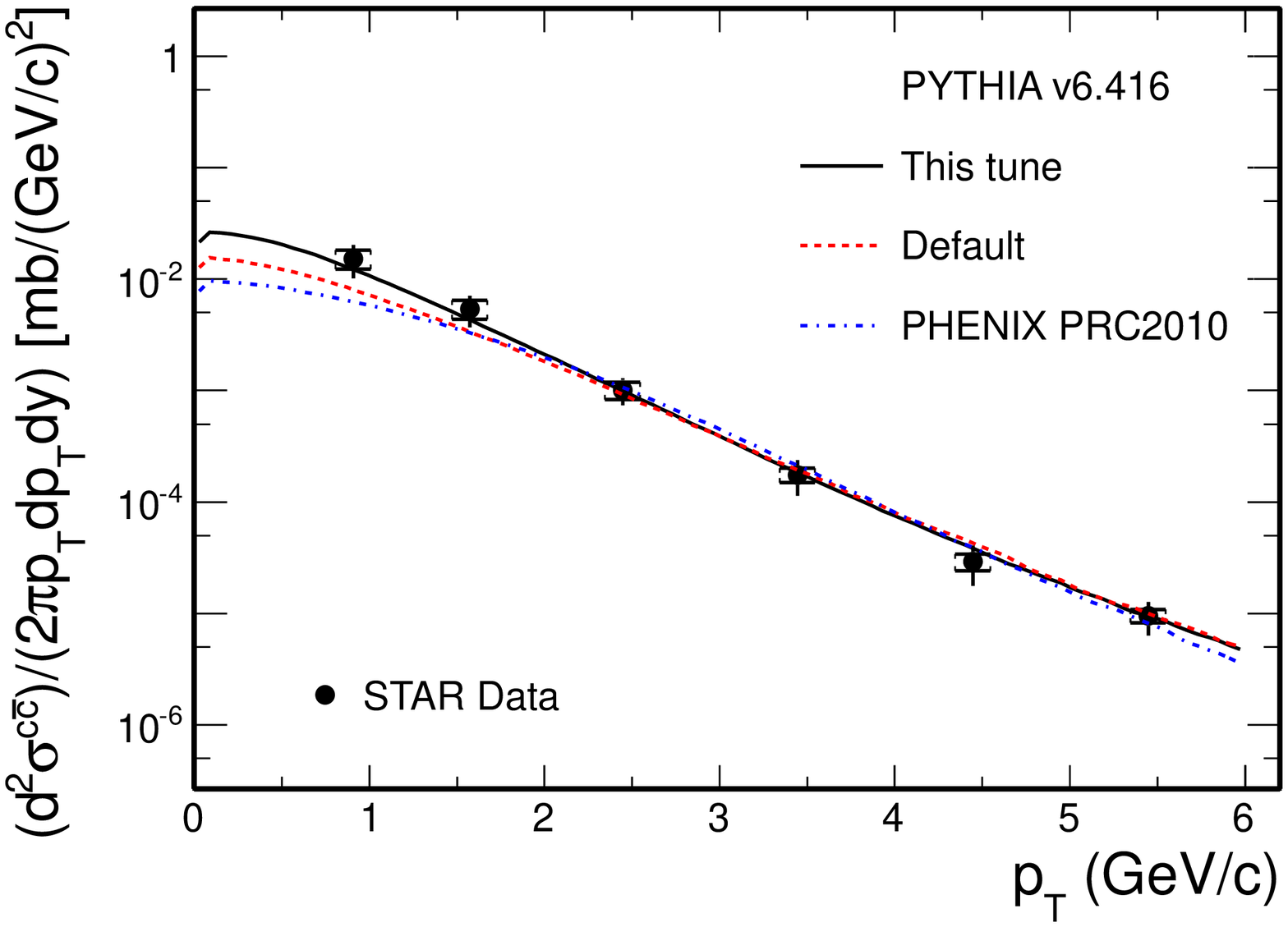}}
\caption[]{(color online) $c\bar{c}$ production cross section as
  inferred from \Dzero and \Dstar production in \pp collisions at \srt
  = 200 GeV compared with PYTHIA calculations. Data are fitted with PYTHIA spectra with a overall scale parameter for the purpose of shape comparison only.} 
\label{fig15PYTHIAspectra}
\end{figure}

Shown in Fig.~\ref{fig15PYTHIAspectra}, the data were also compared
with PYTHIA calculations. PYTHIA version 6.416 was used as it has been
tuned to describe the mid-rapidity Tevatron data. We tried PYTHIA calculations with the
following sets of parameters to compare with our measurements:

\begin{itemize}
\item[a)] Default MSEL = 1.
\item[b)] PHENIX tune: MSEL = 0 with MSUB(11,12, 13, 28, 53, 68) on, PARP(91) ($\langle k_{\perp}\rangle$) = 1.5 GeV/$c$, MSTP(32) ($Q^2$ scale) = 4, CKIN(3) (min. parton $\hat{p_{\perp}}$) = 2 GeV.
\item[c)] This tune: MSEL = 1, PARP(91) ($\langle k_{\perp}\rangle$) = 1.0 GeV/$c$, PARP(67) (parton shower level) = 1.0.
\end{itemize}

The choice of modifying the primordial $\langle k_{\perp}\rangle$ (the Gaussian width of primordial $k_T$
in hadrons) and
the parton shower level parameters from default values (2 GeV/$c$ and
4, respectively) in this tune was suggested by the matching of scales
in heavy-flavor production at lower energies~\cite{HFPYTHIA}, which
has been noted in PYTHIA~\cite{pythia}. The CDF tuneA
parameters~\cite{CDF:TuneA}, which were tuned to reproduce mid-rapidity
jet and ``underlying event'' results at Tevatron energies, are included as defaults in PYTHIA
v6.416. ``PHENIX tune'' parameters are those used in the PHENIX charm
continuum contribution estimation from dielectron measurements~\cite{PHENIXdielectron}. The default parton distribution function (CTEQ5L) was used in all three cases.

All ground-state charmed hadrons ($D^0$, $D^+$, $D^+_s$, and
$\Lambda^+_c$) were added together in the rapidity window $|y|<$ 1 to
obtain charm cross sections. 
The data was then fitted with the PYTHIA calculations with a overall scale factor as the unique free parameter. The charm production $p_T$ spectrum with this tune gives best $\chi^2$: 1.41 (this tune), 4.97 (default), 5.96 (PHENIX tune). This is the first direct D-meson measurement that goes down to such a low $p_T$, which constrains the model parameters better. 


\vspace{+0.2cm}

\section{Summary}

In summary, measurement on the charmed meson (\Dzero and $D^{*}$) production cross sections via their hadronic decays in \pp collisions at $\sqrt{s} = $ 200 GeV has been reported. The charm pair production cross section at mid-rapidity extracted from this analysis is \dsigmady = 170 $\pm$ 45 (stat.) $^{+38}_{-59}$ (sys.) $\mu$b. The charm total cross section at \srt = 200\,GeV is estimated as 797 $\pm$ 210 (stat.) $^{+208}_{-295}$ (sys.) $\mu$b. The reconstructed charmed mesons cover the $p_T$ range from 0.6-6 GeV/$c$. The charm-pair transverse momentum differential cross sections from this analysis are consistent with the upper bound of a Fixed-Order Next-to-Leading Logarithm perturbative QCD calculation. When comparing to PYTHIA model calculations, we found that a calculation with smaller primordial $\langle k_{\perp}\rangle$ and parton shower level compared to CDF TuneA settings describes the shape of the $p_T$ distribution of data. 


\section{Acknowledgement}
We thank the RHIC Operations Group and RCF at BNL, the NERSC Center at
LBNL and the Open Science Grid consortium for providing resources and
support. This work was supported in part by the Offices of NP and HEP
within the U.S. DOE Office of Science, the U.S. NSF, the Sloan
Foundation, the DFG cluster of excellence `Origin and Structure of the
Universe' of Germany, CNRS/IN2P3, FAPESP CNPq of Brazil, Ministry of
Ed. and Sci. of the Russian Federation, NNSFC, CAS, MoST, and MoE of
China, GA and MSMT of the Czech Republic, FOM and NWO of the
Netherlands, DAE, DST, and CSIR of India, Polish Ministry of Sci. and
Higher Ed., Korea Research Foundation, Ministry of Sci., Ed. and
Sports of the Rep. of Croatia, and RosAtom of Russia.


\end{document}